\documentclass[12pt]{article}
\usepackage{color}
\usepackage{amsmath}
\usepackage{amsfonts,amssymb} 
\usepackage{graphicx}
\usepackage{gensymb}
\usepackage{times}

\newcommand{\ket}[1]{\vert #1 \rangle }
\newcommand{\bra}[1]{\langle #1 \vert }
\newcommand{\inlinecomment}[1]{}

\newenvironment{sciabstract}{%
\begin{quote} \bf}
{\end{quote}}

\newcounter{lastnote}

\title{Preparation of entangled states through Hilbert space engineering}

\author{Y. Lin,$^{1\dagger\ast}$ J. P. Gaebler,$^{1}$ F. Reiter,$^{2\ddagger}$ T. R. Tan,$^{1}$ \\
R. Bowler,$^{1\mathsection}$ Y. Wan,$^{1}$\\
A. Keith,$^{1}$ E. Knill,$^{1}$S. Glancy,$^{1}$ K. Coakley,$^{1}$\\ A. S. S\o rensen,$^{2}$ D. Leibfried,$^{1}$ D. J. Wineland$^{1}$\\
\normalsize{$^{1}$National Institute of Standards and Technology,}\\
\normalsize{325 Broadway, Boulder, Colorado 80305, USA.}\\
\normalsize{$^{2}$The Niels Bohr Institute, University of Copenhagen,}\\
\normalsize{Blegdamsvej 17, DK-2100 Copenhagen \O, Denmark.}\\
\normalsize{$^{\dagger}$Current address: JILA, National Institute of Standards and Technology and}\\
\normalsize{University of Colorado, and Department of Physics,}\\
\normalsize{University of Colorado, Boulder, CO 80309, USA}\\
\normalsize{$^{\ddagger}$Current address: Institute for Quantum Optics and Quantum Information, }\\
\normalsize{Austrian Academy of Sciences, 6020 Innsbruck, Austria}\\
\normalsize{$^{\mathsection}$Current address: University of Washington, }\\ \normalsize{Department of Physics, Box 351560, Seattle, Washington 98195, USA. }\\
\normalsize{$^\ast$To whom correspondence should be addressed; E-mail:  yiheng.lin@colorado.edu}}

\date{}

\begin{document}

\baselineskip24pt


\maketitle

\begin{sciabstract}
Entangled states are a crucial resource for quantum-based technologies such as  quantum computers and quantum communication systems \cite{Nielsen2011,Gisin2002}. Exploring new methods for entanglement generation is important for diversifying and eventually improving current approaches. Here, we create entanglement in atomic ions by applying laser fields to constrain the evolution to a restricted number of states, in an approach that has become known as ``quantum Zeno dynamics'' \cite{Frerichs1991,Facchi2002,Facchi2008}. With two trapped $^9\rm{Be}^+$ ions, we obtain Bell state fidelities up to $0.990^{+2}_{-5}$; with three ions, a W-state \cite{Duer2000} fidelity of $0.910^{+4}_{-7}$ is obtained. Compared to other methods of producing entanglement in trapped ions, this procedure is relatively insensitive to certain imperfections such as fluctuations in laser intensity, laser frequency, and ion-motion frequencies.
\end{sciabstract}


The quantum Zeno effect usually refers to the inhibition of quantum dynamics due to frequent measurements \cite{Misra1977,Itano1990,Balzer2002}. More generally, the idea is to restrict the dynamics to a subspace of the overall system. Recent proposals \cite{Maniscalco2008,Wang2008,Raimond2010,Smerzi2012,Burgarth2013,Li2013,Zhu2014,Zhang2015} have explored ways to provide this subspace isolation by coupling the remainder of the system to auxiliary quantum states. This situation has become known as quantum Zeno dynamics, though the restrictions can be implemented by unitary interactions without the need for measurements. Dynamics in a restricted subspace have recently been demonstrated with atoms in Bose-Einstein condensates \cite{Schaefer2014}, Rydberg atoms \cite{Signoles2014,Jau2015a}, atoms in a cavity \cite{Barontini2015} and photons in a cavity coupled to a superconducting qubit \cite{Bretheau2015}. Here, we apply coherent laser fields to trapped ions to confine their quantum evolution to an effective two-level subspace consisting of an initial product state and an entangled state. With this Hilbert space engineering, we prepare an entangled state by applying a spatially uniform microwave control pulse to a collection of ions initially in a separable state. This technique can produce high fidelity entangled states with resilience to technical laser noise and fluctuations of the frequencies of ion motion. For measurement-based Zeno dynamics, there is a finite probability of irretrievably escaping from the desired subspace. However, if the subspace restriction is brought about by coherent interactions, the evolution is ideally unitary, and thus state amplitudes that leak from the restricted subspace remain coherent and can be recovered with additional coherent operations. We demonstrate this advantage of coherent subspace engineering by applying a composite pulse sequence, and observe an improved fidelity of the entangled state.

\begin{figure}[!ht]
  \centering
\includegraphics[scale=0.5]{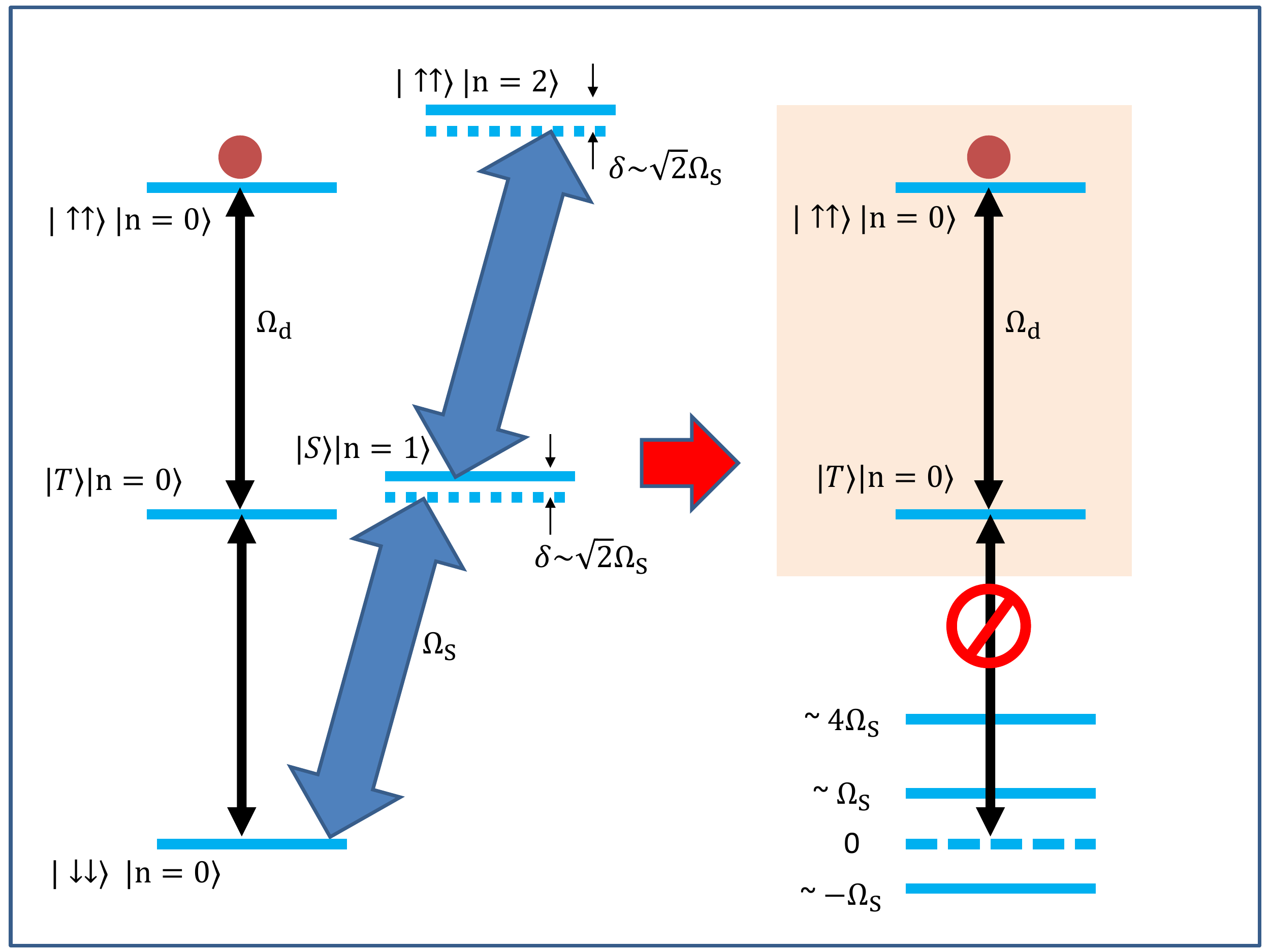}
  \caption{Restricted dynamics for two ions. The thin black arrows depict the relatively weak microwave coupling; the thick blue arrows depict laser-induced strong blue sideband coupling. With the $\ket{{\uparrow\uparrow}}$ state initially populated (red dots), in the absence of the sideband excitation, the microwaves drive the state down the symmetric manifold (the states on the left) with Rabi frequency $\Omega_d$, where the $\ket{T}$ and $\ket{S}$ states are defined in the text, and such a global rotation alone cannot generate entanglement. However, the sideband excitations (with Rabi frequency $\Omega_s$) dress the $\ket{{\downarrow\downarrow}}$ state, shifting its components out of resonance with respect to the weak microwave drive, as shown on the right. Thus given $\Omega_s\gg\Omega_d$, the microwave drive only couples the two highest energy states in the symmetric manifold, and the entangled $\ket{T}$ state can be created with an effective $\pi$ pulse of the microwave drive ($t_{\pi}=\pi / (2 \sqrt{2} \Omega_d)$) from the $\ket{{\uparrow\uparrow}}$ state. }\label{fig:Dicke}
\end{figure}

When applying a global rotation to an initial state with $N$ two-level (spin-$\frac{1}{2}$) systems in the spin up state $\ket{{\uparrow}}$, each spin rotates independently and the overall quantum state remains separable. The evolution can be described in the symmetric angular momentum manifold $\ket{J=N/2,m_J}$ \cite{Arecchi1972}, or Dicke states \cite{Dicke1954}, where $J$ is the total angular momentum quantum number and $m_J$ is the projection of the angular momentum along the quantization axis. All individual $\ket{J,m_J}$ states are entangled states except the maximal spin states, $\ket{{\uparrow\uparrow...\uparrow}} = \ket{J,J}$ and $\ket{{\downarrow\downarrow...\downarrow}} = \ket{J,-J}$. Entanglement between multiple spins can be generated by perturbing specific $\ket{J,m_J}$ states in the manifold to restrict the dynamics. A simple case is to apply a perturbation to shift the $\ket{J,J-2}$ state out of resonance, as depicted in Fig. \ref{fig:Dicke} for the case of two spins. In this case, the dynamics are restricted within the $\ket{J,J}$ and $\ket{J,J-1}$ states. Thus, starting from $\ket{J,J}$, the entangled $\ket{J,J-1}$ state \cite{Zeilinger1992,Duer2000} is prepared by an effective $\pi$-pulse. For two and three spins, these states are the triplet Bell state $\ket{T}=\frac{1}{\sqrt{2}}(\ket{{\uparrow\downarrow}}+\ket{{\downarrow\uparrow}})$ and the W-state \cite{Duer2000} $\ket{W}=\frac{1}{\sqrt{3}}(\ket{{\uparrow\uparrow\downarrow}}+\ket{{\uparrow\downarrow\uparrow}}+\ket{{\downarrow\uparrow\uparrow}})$, respectively.

We experimentally demonstrate this scheme with trapped $^9\rm{Be}^+$ ions aligned along the axis of a linear Paul trap \cite{Drees1964,Raizen1992,Blakestad2009}. In an applied magnetic field of 11.946 mT, the frequency splitting $\omega_0\approx2\pi\times 1.2075$ GHz between the $^2S_{1/2}$ hyperfine ground states $\ket{F=2,m_F=0}\equiv\ket{{\downarrow}}$ and $\ket{F=1,m_F=1}\equiv\ket{{\uparrow}}$ is first-order insensitive to magnetic field fluctuations \cite{Langer2005}. The effective rotation in the restricted subspace is produced by a uniform resonant microwave field, while the restricting perturbations are provided by a laser-induced coupling between ions via a shared motional mode.
With two ions and without applied laser fields, the microwave field couples the Dicke states with the Hamiltonian
\begin{equation}
H_d = \hbar \Omega_d\sum_{i=1,2}\sigma_i^x = \sqrt{2}\hbar \Omega_d(\ket{{\downarrow\downarrow}}\bra{T} + \ket{T}\bra{\uparrow\uparrow}) + {\rm H.c.},
\end{equation}
where $\hbar $ is the reduced Planck constant, $\Omega_d$ is the single-ion Rabi frequency, $\sigma_i^x$ is the Pauli operator on the $i^{\rm{th}}$ ion, and $\rm H.c.$ stands for Hermitian conjugate. If the spins are initially in a product state, evolution under this Hamiltonian will not generate entanglement.

To generate the desired dynamics for two ions, we address the ``stretch'' axial normal mode of motion of frequency $\omega\approx2\pi \times 6.20$ MHz, with a laser-induced stimulated-Raman blue sideband interaction \cite{SUPP}. The sideband interaction is detuned from resonance by $\delta$, and is described by the Hamiltonian
\begin{equation}
H_s = \hbar \Omega_s (\sigma_1^- - \sigma_2^-)a e^{-i \delta t} + {\rm H.c.},\label{eq:H_s}
\end{equation}
where $\Omega_s$ is the Rabi frequency, $a$ is the annihilation operator of the stretch mode, and $\sigma_i^-=\ket{{\downarrow}}_i\bra{{\uparrow}}$ is the spin lowering operator for ion $i$. In Eq. (\ref{eq:H_s}), we have assumed that the Raman phase on the two ions is the same (modulo $2\pi$). The minus sign between the two spin lowering operators results from the stretch-mode amplitudes being equal but opposite for the two ions. The symmetry of the $\ket{{T,n}}$ state implies that the sideband interaction does not couple this state to other relevant states. However, as depicted on the left in Fig. \ref{fig:Dicke}, it couples the states $\ket{{\downarrow\downarrow}}\ket{n}\leftrightarrow\ket{{S}}\ket{n+1}\leftrightarrow\ket{{\uparrow\uparrow}}\ket{n+2}$, where $\ket{n}$ denotes a stretch mode Fock state, and $\ket{S}=\frac{1}{\sqrt{2}}(\ket{{\uparrow\downarrow}}-\ket{{\downarrow\uparrow}})$. The energies of the resulting dressed states (the eigenstates of the ions with $H_s$ included) are shifted to approximately $\pm\hbar\Omega_s$ and $4\hbar\Omega_s$ (right hand side of Fig. \ref{fig:Dicke}), when the detuning $\delta$ is set to approximately $\sqrt{2}\Omega_s$ \cite{SUPP}, so that the energy shift can be made large compared to $\hbar\Omega_d$ for $\Omega_s\gg\Omega_d$. In addition, $H_s$ couples $\ket{{\uparrow\uparrow,n}}$ to $\ket{{S,n-1}}$ for $n>0$, but these couplings are absent if we initialize the stretch mode in the ground state $n=0$. If $\Omega_s\gg\Omega_d$, the system evolves as an effective two-level system between $\ket{{\uparrow\uparrow}}\ket{0}$ and $\ket{{T}}\ket{0}$ under the combined influence of $H_s$ and $H_d$, within a subspace isolated from other states. This allows the preparation of the entangled state $\ket{{T}}\ket{0}$ by a single effective $\pi-$pulse from $\ket{{\uparrow\uparrow}}\ket{0}$. However, for $n > 0$, the desired subspace will not be isolated; therefore, high fidelity motional ground state preparation is crucial \cite{SUPP}.

To initialize the spin and motional states, we first sideband cool both axial modes of the ions to near the ground state, achieving average motional occupation of $\bar{n}<0.006$ for the stretch mode \cite{Monroe1995}. Optical pumping prepares both ions in the $\ket{F=2,m_F=2}$ atomic state. We then apply a global composite microwave $\pi$-pulse to initialize to the $\ket{{\uparrow\uparrow}}$ state \cite{SUPP,Levitt1986}. We set the laser beam and microwave intensities to give $\Omega_s \approx 2\pi\times 17.6$ kHz and $\Omega_d \approx  2\pi\times 1.52$ kHz. We choose $\delta \approx  2\pi\times 27.1$ kHz while maintaining a Raman detuning of approximately 480 GHz red detuned from the $^2P_{1/2}$ state. We simultaneously apply microwaves and laser beams for a variable duration $t$, followed by detection pulses. We observe coherent Rabi flopping between the $\ket{{\uparrow\uparrow}}$ and $\ket{T}$ states as shown in Fig. 2, where the population in the $\ket{{\uparrow\uparrow}}$ and $\ket{{\downarrow\downarrow}}$ states, and the fidelity of the $\ket{T}$ state are determined as described in the supplementary material.


\begin{figure}[!ht]
  \centering
\includegraphics[scale=0.35]{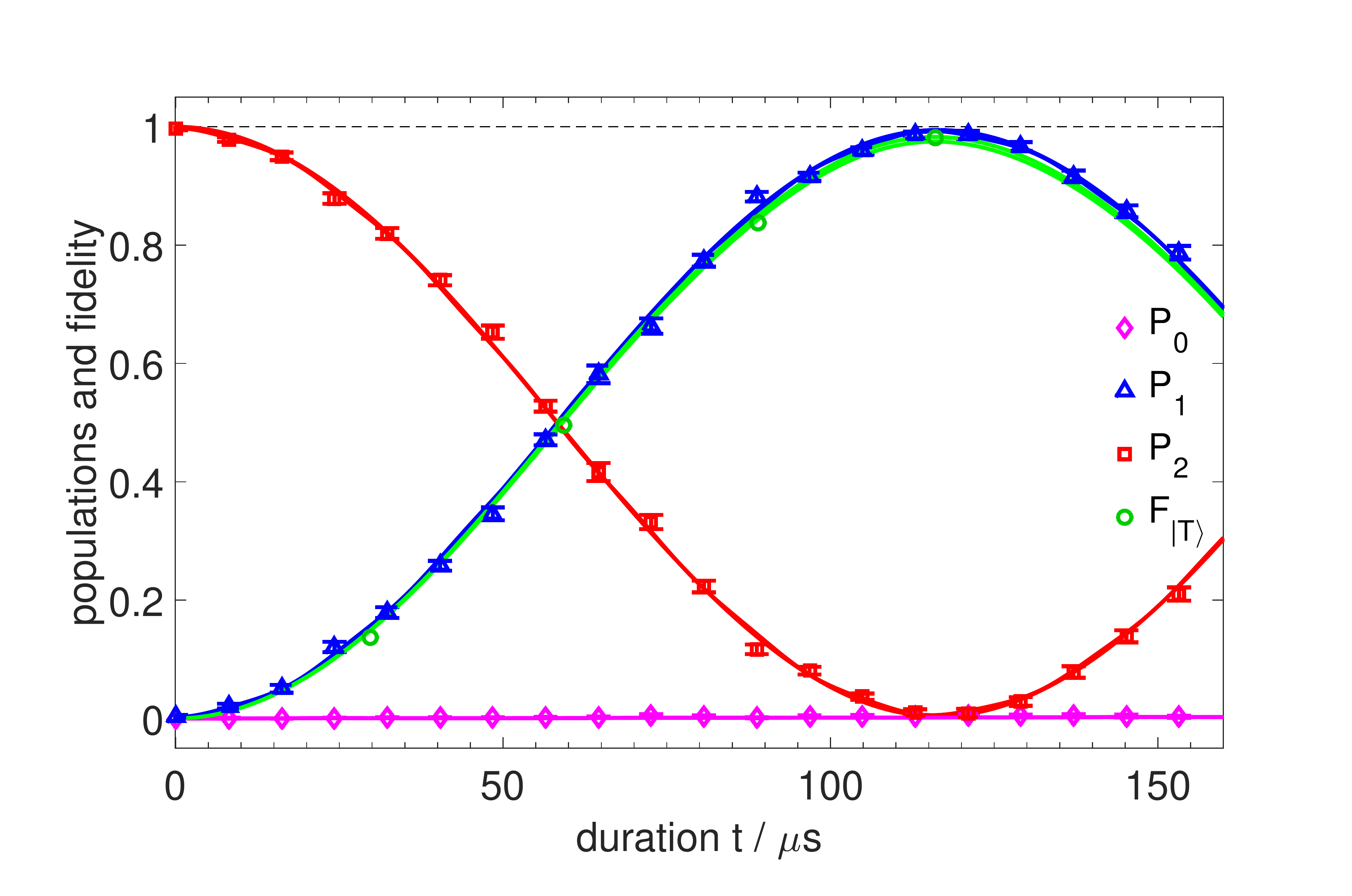}
  \caption{Two-ion population evolution and $\ket{{T}}$ state fidelity for restricted dynamics with microwave and sideband excitations applied simultaneously. Population mainly evolves between the $\ket{{\uparrow\uparrow}}$ and the $\ket{T}$ state, while other states have very small populations. The black dashed line shows unit population/fidelity. The pink diamonds, blue triangles, red squares and green circles represent the measured populations of states with no spins up $P_0$, one spin up $P_1$, two spins up $P_2$, and the fidelity of the $\ket{T}$ state $F_{\ket{{T}}}$, respectively. The population measurements are obtained by repeating the experiment 1,500 times; the fidelity points are derived from 60,000 experiments \cite{SUPP}.
  The difference between $P_1$ and $F_{\ket{{T}}}$ is due to the population in the $\ket{{S}}$ state. The solid lines show the results of a numerical simulation taking into account all known experimental imperfections, with the same coloring convention as for the measured populations. We run the simulation with and without including an upper bound on the imperfections of cooling and spin state initialization. The results of these two simulations are indistinguishable on the scale shown in the figure. The populations and fidelity are inferred by means of a maximum likelihood analysis and the error bars represent the uncertainties according to parametric bootstrap resampling \cite{SUPP}. The uncertainties of $F_{\ket{{T}}}$ are smaller than the symbols. }\label{fig:twoion}
\end{figure}


We observe a maximal fidelity of the $\ket{T}$ state of $0.981^{+2}_{-4}$  after a duration of $t_\pi\approx$ 116 $\rm{\mu}$s, which matches the theoretical prediction \cite{SUPP} of $t_{\pi}=\pi / (2 \sqrt{2} \Omega_d)$. The fidelities and error bars are derived from maximum likelihood partial state tomography, parametric bootstrap resampling, and estimation of state preparation errors \cite{SUPP}. The largest error contributions are estimated to be 0.010 from insufficient isolation of the subspace ($\Omega_s/ \Omega_d\approx$ 12), 0.008 from spontaneous emission \cite{Ozeri2007}, less than 0.006 from imperfect ground state cooling, and less than 0.002 from imperfect initialization of the $\ket{{\uparrow\uparrow}}$ spin state \cite{SUPP}. We compare our data to a numerical simulation including these errors (solid lines in Fig. 2) and find good agreement.



The evolution is ideally unitary; therefore neglecting spontaneous emission and heating of the motional normal mode, any state amplitudes outside the desired subspace can be recovered. To demonstrate this, we apply a specifically tailored composite pulse pair which enables us to return the population in the undesired states $\ket{{\downarrow \downarrow , n=0}} $, $\ket{S,n=1}$, and $\ket{{\uparrow \uparrow, n=2}}$ into the isolated subspace and thereby increase the population of $\ket{T}$. To do this we split the laser pulse into two segments of duration $t_1$ and $t_2$, changing the laser phase by $\pi$ and the sideband detuning from $\delta_1$ to $\delta_2=-\delta_1$. States outside the desired subspace are driven nonresonantly from the $\ket{T}$ state. The amplitudes of these undesired states get a contribution from each of the two pulse segments, leading to an interference between the two contributions, reminiscent of the two-pulse interference in Ramsey spectroscopy. Within first order perturbation theory one can show that the amplitudes of all undesired states interfere destructively and vanish at the time where the fidelity of $\ket{T}$ is maximal if one sets $\delta_1 = -\delta_2 = \sqrt{7/3} \Omega_s$, $\Omega_d = \Omega_s / (3 \sqrt{6})$, and $t_2=2t_1$. When the amplitudes of the undesired states vanish, the associated constructive interference is in the amplitude of the $\ket{T}$ state which will have a near unity population only limited by higher order effects \cite{SUPP}. Experimentally we set $\Omega_s = 2\pi\times 17.3$ kHz, $\Omega_d = 2\pi\times 2.55$ kHz, $\delta_1 = -\delta_2 = 2\pi\times 26.8$ kHz, $t_1$ = 25.4 $\rm{\mu s}$, and $t_2$ = 47.3 $\rm{\mu s}$ to obtain a $\ket{T}$ state population of $0.990^{+2}_{-5}$. The symbols in Fig. 3 show the experimentally observed population evolution during the composite pulse sequence, in agreement with numerical simulations (solid lines). Higher fidelity is achieved despite a smaller ratio $\Omega_s / \Omega_d \approx 7$, by recovering amplitudes that leaked out due to insufficient isolation of the subspace, reducing this error to 0.001 (We note that according to simulations, further reduction can be achieved with better calibration of $t_1$). The reduced $\Omega_s / \Omega_d$ has the beneficial effect of suppressing the spontaneous emission error to 0.005. Similar to the single-pulse experiment, we estimate errors less than 0.005 from imperfect ground state cooling, and less than 0.002 from imperfect initialization of the $\ket{{\uparrow\uparrow}}$ spin state \cite{SUPP}. We compare our data to a numerical simulation including these errors (solid lines in Fig. 3) and find good agreement.

\begin{figure}[!ht]
  \centering
\includegraphics[scale=0.35]{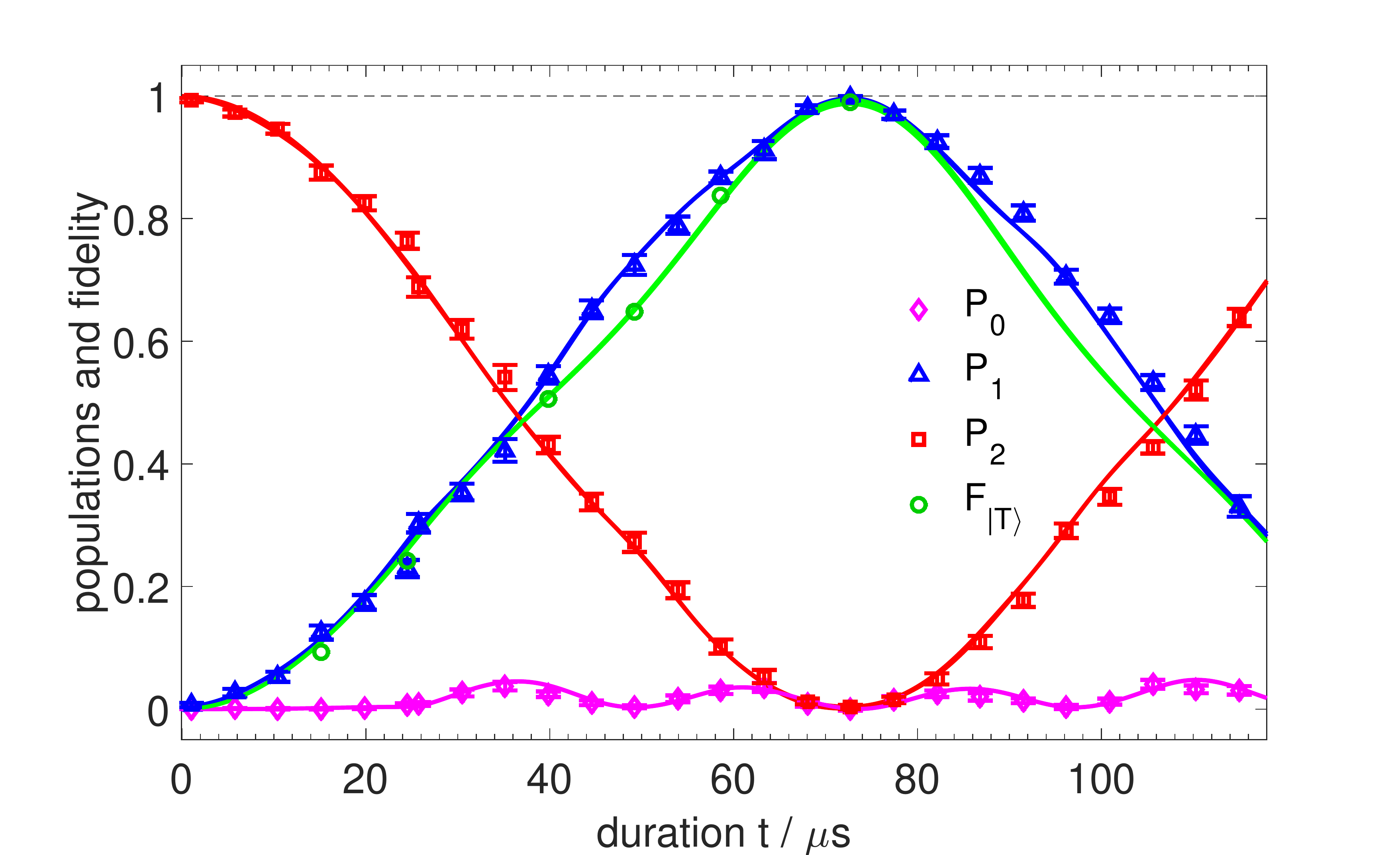}
  \caption{State evolution for restricted dynamics of two trapped ions using a composite pulse sequence. Similar to Fig. 2, populations are mainly confined to the $\ket{{\uparrow\uparrow}}$ and $\ket{T}$ states. The coloring and labeling conventions are the same as in Fig. 2. The laser beam phase and detuning are flipped 25.4 $\rm{\mu s}$ after the start of the experiment. Note that the oscillations of $\ket{{\downarrow\downarrow}}$ are enhanced for $t>$ 25.4 $\rm{\mu s}$; however the maximal population of the $\ket{T}$ state is increased compared to the single pulse used for the data in Fig. 2. We numerically simulate this experiment with and without including an upper bound of imperfections of cooling and spin state initialization. The simulation results overlap on the scale of the figure. The populations, fidelity and error bars are inferred as in Fig. 2 \cite{SUPP}. The population measurements are obtained by repeating the experiment 1,000 times; the fidelity points are derived from 40,000 experiments \cite{SUPP}. The uncertainties of $F_{\ket{{T}}}$ are smaller than the symbols.}\label{fig:twopulse}
\end{figure}



We also demonstrate restricted dynamics on three $^9\rm{Be}^+$ ions. We tune the laser beam frequencies to address the center-of-mass (COM) mode blue sideband, which has equal mode amplitudes on each ion. The ion spacings are set such that the phase of the sideband interaction on each ion differs by $2\pi/3$ so that the $\ket{W,n=0}$ state will be a dark state of the sideband interaction \cite{SUPP}. Starting from the $\ket{{\uparrow\uparrow\uparrow, n=0}}$ state, and with driving field parameters similar to the case of two ions, we observe flopping between the $\ket{{\uparrow\uparrow\uparrow}}$ and $\ket{W}$ states, in agreement with the numerical simulations \cite{SUPP}. We obtain a $\ket{W}$ state fidelity of $0.910^{+4}_{-7}$ after a duration of 114.1 $\rm{\mu s}$. The sources of infidelity include those of the two-ion case (in general leading to larger imperfections) plus two sizable additions: 0.011 from heating of the COM mode caused by electric field noise and 0.023 from unequal laser illumination on the three ions due to the Gaussian profile of the laser beam \cite{SUPP}.

For more than three ions in a chain, numerical simulations and analytic analysis indicate the presence of unwanted dark states such that straightforward application of the sideband interaction does not yield an effective two-level system between the first two Dicke states. However, by using a combination of sideband laser interactions on multiple motional modes and engineering the relative phases of the sideband couplings on each ion, the scheme may be scaled up to isolate an effective two-level system of multiple spins \cite{Reiter2015}.

In summary, we describe and demonstrate a scheme to isolate subspaces of spin states with trapped ions, enabling the creation of entangled states by the application of global uniform oscillating fields. We create a two-ion triplet Bell state with fidelity of $0.990^{+2}_{-5}$, and a three-ion $\ket{W}$ state with fidelity of $0.910^{+4}_{-7}$. The entangled state fidelity is relatively insensitive to fluctuations in laser power and frequency and motional mode frequency fluctuations, since the main requirement is that the frequency shifts due to the laser-induced spin-motion coupling are large compared to the microwave Rabi frequency, but the exact value and stability of the shifts are not crucial. Therefore, this scheme may serve as an alternative way of preparing entangled states, without using conventional multi-qubit entangling quantum logic gates \cite{Haeffner2005}. This work also presents an application of Hilbert space engineering, which may be extended to generate other entangled states or spin dynamics. Our scheme can be generalized to other experimental platforms, for example superconducting qubits or atoms in a cavity. 


We thank J. Eschner, G. Morigi, and A. Signoles for helpful discussions. We thank C. Kurz and D. H. Slichter for helpful comments on the manuscript. The NIST work was supported by the Office of the Director of National Intelligence (ODNI) Intelligence Advanced Research Projects Activity (IARPA), ONR, and the NIST Quantum Information Program.  The work of the University of Copenhangen was funded by the European Union Seventh Framework Programme through ERC Grant QIOS (Grant No. 306576).  This paper is a contribution by NIST and not subject to U.S. copyright.


\clearpage

\renewcommand{\figurename}{\textbf{Fig.}}
\renewcommand\thefigure{\textbf{S\arabic{figure}}}
\renewcommand{\tablename}{\textbf{Table}}
\renewcommand\thetable{\textbf{S\arabic{table}}}

\section*{Supplementary Material}
\setcounter{figure}{0}

\section{Motional modes for two and three ion chains}
~~~~~For two ions, the axial modes of motion are the center-of-mass ``COM'' mode and the stretch mode. In our experiment, these frequencies were approximately $2\pi \times \{3.58, 6.20\}$ MHz. The COM and stretch normal mode amplitudes for the two ions are $\{\{\frac{1}{\sqrt{2}},\frac{1}{\sqrt{2}}\},\{\frac{1}{\sqrt{2}},-\frac{1}{\sqrt{2}}\}\}$ respectively (the two ions oscillate in phase and out-of-phase for these motional modes respectively). We alternately apply red sideband and re-pumping pulses 30 times each to cool the COM and stretch modes to reach motional states that are very close to the asymptotic equilibrium motional occupation \cite{Monroe1995}; the COM and stretch mode mean phonon occupation numbers are determined to be smaller than \{$0.01,0.006$\}, respectively.

For three $^9\rm{Be}^+$ ions in a linear chain, the three axial modes are the COM, stretch, and ``Egyptian'' with frequencies approximately $2\pi\times\{3.60, 6.24, 8.68\}$ MHz, and mode amplitudes \{\{$\frac{1}{\sqrt{3}}$,$\frac{1}{\sqrt{3}}$,$\frac{1}{\sqrt{3}}$\}, \{$\frac{1}{\sqrt{2}}$,0,-$\frac{1}{\sqrt{2}}$\}, \{-$\frac{1}{\sqrt{6}}$,$\frac{2}{\sqrt{6}}$,-$\frac{1}{\sqrt{6}}$\}\} respectively. Following sideband cooling, we determine the COM mode mean phonon occupation number to be approximately 0.02.

\section{Sideband and microwave interactions}

~~~~~To induce the sideband interaction, we apply a pair of laser beams such that the difference of their momentum vectors $\Delta\textbf{k}$ at the site of the ions is aligned along the trap axis and their frequency difference is set to $\omega_0+\omega+\delta$, detuned from the blue sideband of a normal mode of frequency $\omega$ (the stretch mode for two and the COM mode for three ions) by $\delta \ll \{\omega_0,~\omega\}$. For two ions, taking the equilibrium position of ion 1 to be the origin of the axis, we denote the equilibrium position of ion 2 to be $X$  (here $X$  is a number, not an operator). The lasers induce a near-resonant ``blue sideband'' coupling on the stretch mode described in the interaction frame by \cite{Wineland1998}
\begin{equation}
H_s = \Omega_s (\sigma_1^- - e^{i|\Delta\textbf{k}| X} \sigma_2^-)a e^{-i \delta t+i \phi} + {\rm H.c.},\label{eq:H_s_deltak}\tag{S1}
\end{equation}
where the $e^{i|\Delta\textbf{k}| X}$ term represents the differential optical laser phase between the two ions. The minus sign between the $\sigma$ operators results from the opposite motional mode amplitudes of the ions in the stretch mode. The phase difference of the Raman laser beams at the origin for $t=0$ is denoted as $\phi$. To obtain the coupling of Eq. (\ref{eq:H_s}), we adjust the axial confinement such that the ion-spacing $X$ is as close as possible to $M \frac{2\pi}{|\Delta\textbf{k}|}$, with $M$ being an integer number; in our experiment $M=18$. This gives $H_s$ the form quoted in the main text and leads to the desired couplings isolating the subspace, as discussed there and in Sec. 4 below.

For three ions, we apply laser beams tuned close to the blue sideband of the COM mode so that the ions have identical motional amplitudes. We take the center ion equilibrium position to be zero, and the outer ions' equilibrium positions are $\pm X^\prime$. Thus the sideband interaction can be expressed as
\begin{equation}
H_s^\prime  = \Omega_s^\prime  (e^{i |\Delta\textbf{k}| X^\prime}\sigma_1^- + \sigma_2^- + e^{ -i |\Delta\textbf{k}| X^\prime}\sigma_3^-)a e^{-i \delta^\prime  t} + {\rm H.c.} ,\label{eq:H_s_prime_k}\tag{S2}
\end{equation}
where $\delta^\prime$ is the detuning from the sideband resonance. We adjust the inter-ion spacing $X^\prime$ to be as close as possible to $M^\prime \frac{2\pi}{|\Delta\textbf{k}|}$, in our experiment $M^\prime=\frac{46}{3}$. With this we obtain a sideband interaction
\begin{equation}
H_s^\prime  = \Omega_s^\prime  (e^{i 2\pi/3}\sigma_1^- + \sigma_2^- + e^{-i 2\pi/3}\sigma_3^-)a e^{-i \delta^\prime  t + i\phi} + {\rm H.c.} .\label{eq:H_s_prime}\tag{S3}
\end{equation} The microwave coupling has no significant phase difference between the ions and can be described by the Hamiltonian
\begin{align} \label{eq:H_d_prime}\tag{S4}
\begin{split}
H_d^\prime & = \Omega_d^\prime \sum_{i=1,2,3}\sigma_i^x  = \Omega_d^\prime(\sqrt{3}\ket{{\uparrow\uparrow\uparrow}}\bra{W} + 2\ket{W}\bra{\overline{W}} + \sqrt{3}\ket{\overline{W}}\bra{{\downarrow\downarrow\downarrow}}\\
&   ~~~~~~~~~~~~~~~~~~~~~~~~~~~~~~~~-\ket{W_{\rm{ac}}}\bra{\overline{W_{\rm{ac}}}}-\ket{W_{\rm{c}}}\bra{\overline{W_{\rm{c}}}}) + {\rm H.c.},
\end{split}
\end{align}
where $\Omega_d^\prime $ is the Rabi frequency,  $\ket{\overline{W}}=\frac{\ket{{\uparrow\downarrow\downarrow}}+\ket{{\downarrow\uparrow\downarrow}}+\ket{{\downarrow\downarrow\uparrow}}}{\sqrt{3}}$; $\ket{W_{\rm{c}}}=\frac{e^{i 2\pi/3}\ket{{\uparrow\uparrow\downarrow}}+ \ket{{\uparrow\downarrow\uparrow}}+ e^{-i 2\pi/3}\ket{{\downarrow\uparrow\uparrow}}}{\sqrt{3}}$, $\ket{W_{\rm{ac}}}=\frac{e^{-i 2\pi/3}\ket{{\uparrow\uparrow\downarrow}}+ \ket{{\uparrow\downarrow\uparrow}}+ e^{i 2\pi/3}\ket{{\downarrow\uparrow\uparrow}}}{\sqrt{3}}$, $\ket{\overline{W_{\rm{c}}}}=\frac{e^{i 2\pi/3}\ket{{\downarrow\downarrow\uparrow}}+ \ket{{\downarrow\uparrow\downarrow}}+ e^{-i 2\pi/3}\ket{{\uparrow\downarrow\downarrow}}}{\sqrt{3}}$, and \\ $\ket{\overline{W_{\rm{ac}}}}=\frac{e^{-i 2\pi/3}\ket{{\downarrow\downarrow\uparrow}}+ \ket{{\downarrow\uparrow\downarrow}}+ e^{i 2\pi/3}\ket{{\uparrow\downarrow\downarrow}}}{\sqrt{3}}$.

\begin{figure*}[!ht]
  \centering
\includegraphics[scale=0.6]{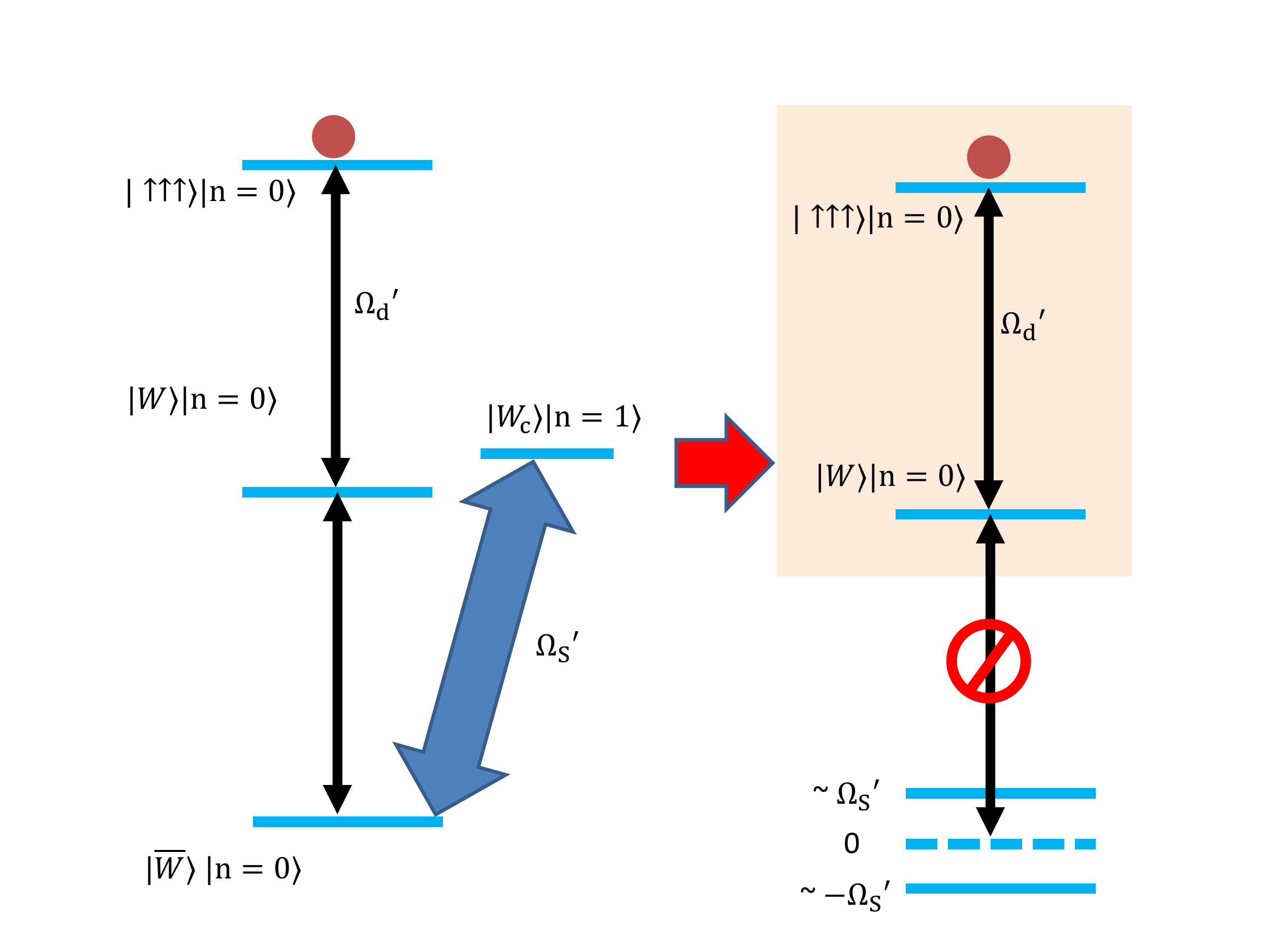}
  \caption{Restricted dynamics for three ions. The thin black arrows depict the relatively weak microwave coupling, and the thick blue arrows depict laser-induced strong blue sideband coupling. With the $\ket{{\uparrow\uparrow\uparrow}}$ state initially populated (red dots), in the absence of the sideband excitation, the microwaves drive the state down the symmetric manifold (the states on the left), where $\ket{W}$ and $\ket{\overline{W}}$ are defined in the text. Such a global rotation alone can not generate entanglement. However, the sideband excitations perturb the $\ket{\overline{W}}$ state, coupling it to $\ket{W_{\rm{c}}}=\frac{e^{i 2\pi/3}\ket{{\uparrow\uparrow\downarrow}}+ \ket{{\uparrow\downarrow\uparrow}}+ e^{-i 2\pi/3}\ket{{\downarrow\uparrow\uparrow}}}{\sqrt{3}}$ and shifting it out of resonance with respect to the weak microwave drive as shown on the right. Thus the microwave drive couples to only the two highest energy states in the symmetric manifold, and the entangled $\ket{W}$ states can be created with an effective $\pi$ pulse driving from the $\ket{{\uparrow\uparrow\uparrow}}$ state. }\label{fig:Dicke3}
\end{figure*}

As depicted in Fig. \ref{fig:Dicke3}, the sideband interaction does not couple to the $\ket{{\uparrow\uparrow\uparrow}}\ket{n=0}$  state, nor states of the form $\ket{W}\ket{n=0}$ since the three components of $H_s^\prime$ lead to a destructive interference of the couplings from $\ket{W}\ket{n=0}$ to $\ket{{\uparrow\uparrow\uparrow,n=1}}$. However, it shifts the energy of the $\ket{\overline{W},n=0}$ state out of resonance, by its coupling to $\ket{W_{\rm{c}},n=1}$, leading to the desired subspace restrictions. As opposed to the two-ion case, the isolated subspace can be achieved with $H_s^\prime$ as a resonant interaction with detuning $ \delta^\prime=0$, since the $\ket{\overline{W},n=0}$ state only couples to $\ket{W_{\rm{c}}}\ket{n=1}$. Thus with $\Omega_s^\prime \gg \Omega_d^\prime$, and the axial modes in the ground state, the weak microwave drive only couples the initial state to the $\ket{W}$ state, while further coupling to the $\ket{\overline{W}}$ state is off resonance. Similar to the two-ion case described in Sec. 4.2, we can thus create the $\ket{W}$ state with a single microwave effective $\pi$-pulse where the effective $\pi$-time is $\pi/(2\sqrt{3}\Omega_d^\prime)$. In the experiment we set $\Omega_s^\prime = 2\pi\times 19.0 $ kHz and $\Omega_d^\prime = 2\pi\times 1.24$ kHz.

\section{Spin readout}
~~~~~We measure spin populations by transferring the states $\ket{{\uparrow}}$ and $\ket{{\downarrow}}$ to other hyperfine states that are maximally distinguishable with laser-induced resonance fluorescence. We transfer population from the $\ket{{\uparrow}}$ state to the $\ket{F=2, m_F=2}$ state using a composite microwave $\pi$ pulse. The pulse sequence is $\{\frac{\pi}{2},0\}-\{\frac{3\pi}{2},\frac{\pi}{2}\}-\{\frac{\pi}{2},0\}$ \cite{Levitt1986}, where $\{\Theta,\Phi\}$ denotes the rotation angle and the azimuthal angle of a vector in the $x$-$y$ plane of the Bloch sphere about which the spin is rotated. We then use a microwave $\pi$ pulse to transfer population from the $\ket{{\downarrow}}$ states to the $\ket{F=1, m_F=-1}$, and another microwave $\pi$ pulse to transfer any residual $\ket{{\downarrow}}$ state population to the $\ket{F=1, m_F=0}$ state. We finally apply a $\sigma^+$ laser beam resonant with the cycling transition between the $^2S_{1/2}\ket{F=2, m_F=2}$ and $^2P_{3/2}\ket{F=3,m_F=3}$ states and record ion fluorescence counts for 330 $\rm{\mu s}$ on a photo-multiplier tube. For the two (three) ion experiments, we count approximately 39 (37) photons for each ion transferred to the $\ket{{2,2}}$ state, and 3 photons for each ion transferred to the $\ket{{1,-1}}$ or $\ket{{1,0}}$ states, with negligible constant stray light and dark count background.

A straightforward method to extract populations would be to approximate the histograms by sums of Poissonian distributions. However, imperfect polarization and off-resonant transitions in the detection process, such as optical pumping, give rise to deviations from simple Poissonians that could lead to erroneously inferred populations. Pumping effects can be accounted for, if the histograms are analyzed with the maximum likelihood (ML) partial tomography algorithm outlined below. For two-ion experiments, it implicitly infers the probabilities of zero, one and two ions in the $\ket{{F = 2, m_F = 2}}$ state, denoted as $P_i , ~i=\{0,1,2\}$, respectively. Neglecting mapping errors, these probabilities can be assigned to zero to two ions in the $\ket{{\uparrow}}$ state.

To obtain the $\ket{{T}}$ state fidelity, we repeat the experiment and insert a microwave $\{\pi/2, \Phi\}$ ``analysis'' pulse with variable phase $\Phi$ before the state transfer pulses and collection of  fluorescence counts histograms. This pulse rotates the $\ket{{T}}$ state into superpositions of $\ket{{\uparrow\uparrow}}$ and $\ket{{\downarrow\downarrow}}$ states, while the $\ket{{S}}$ state is invariant. All collected histograms, with and without the analysis pulse, and additional reference histograms, which we assume to detect known populations (see below), form the input to the ML algorithms to infer the fidelity of the $|T\rangle$ state, denoted as $F_{\ket{{T}}} $.

\begin{figure*}[!ht]
  \centering
\includegraphics[scale=0.35]{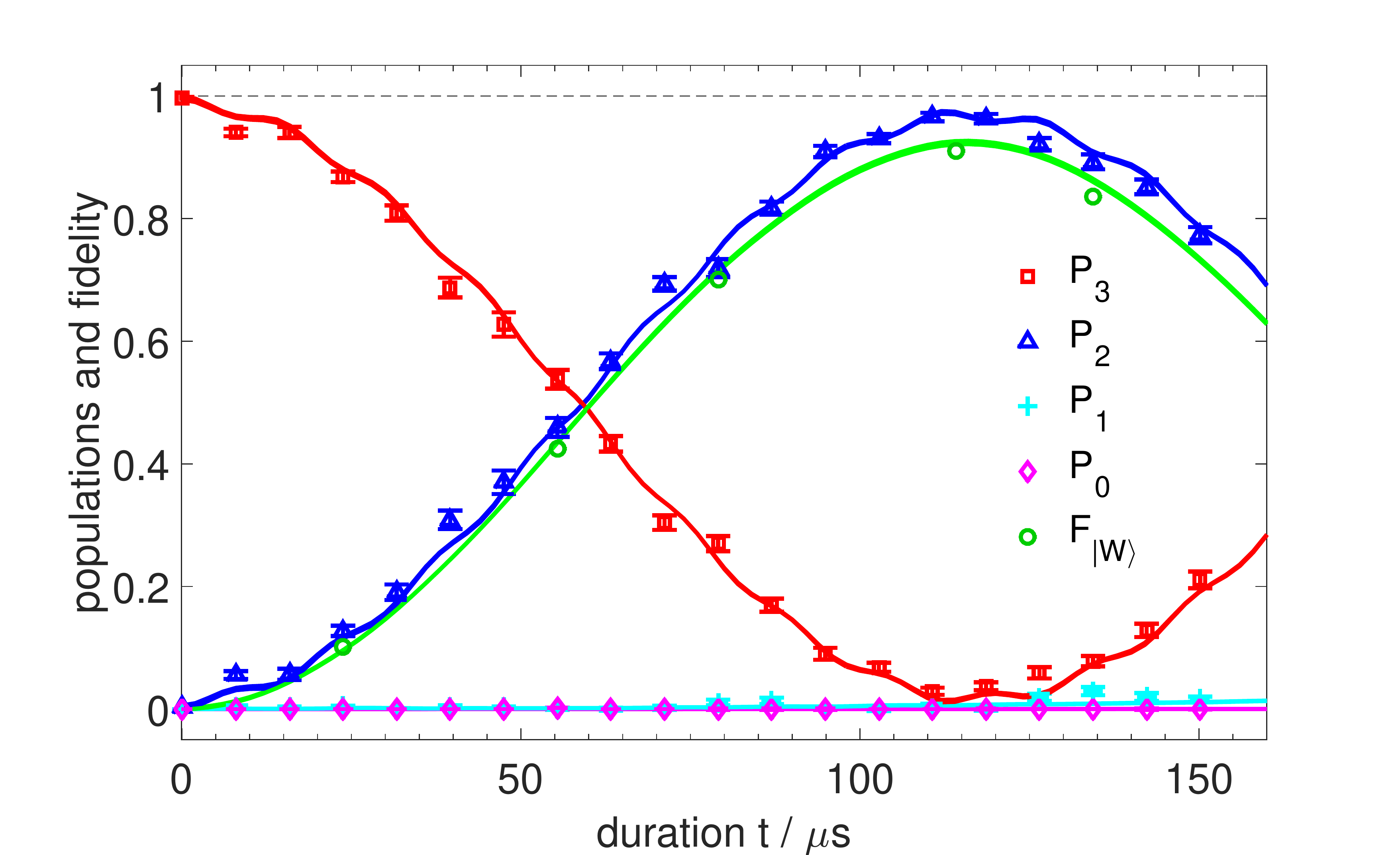}
  \caption{Population evolution for three ions. The red squares, blue triangles, cyan crosses, pink diamonds, and green circles represent the measured probabilities of three spins up, two, one and no spin up and the fidelity of the $\ket{W}$ state,
  denoted as $P_i$ ($i$ = 3-0) and $F_{\ket{{W}}}$, respectively. Solid lines are the result of the numerical simulation, with and without the imperfection of spin state initialization. The simulation results are overlapping on the scale shown in the figure. The population measurements are obtained by repeating the experiment 1,000 times, and for the fidelity measurements we take additional data, as described in the text. The uncertainties of $F_{\ket{{T}}}$ are smaller than the labels.}\label{fig:threeion}
\end{figure*}

For three ions, the detection process is similar to two ions, except the rotation angle of the analysis pulse is $\Theta=\rm{arccos}(1/3)$. Such an operation rotates $\ket{{W}}$ to a superposition of the $\ket{{\uparrow\uparrow\uparrow}}$, $\ket{{\overline{W}}}$ and $\ket{{\downarrow\downarrow\downarrow}}$ states, while the rotation on $\ket{{W_c}}$ and $\ket{{W_{ac}}}$ states retains 2/3 of their populations in the $P_2$ population. Thus we distinguish the $\ket{W}$ state from the $\ket{{W_c}}$ and $\ket{{W_{ac}}}$ states. Provided with these data and the reference histograms, the ML partial tomography algorithm can unambiguously deduce the probabilities that zero to three ions are in the $\ket{{\uparrow}}$ state, denoted as $P_i , i=\{0,1,2,3\}$, respectively and $F_W$, the fidelity of the $|W\rangle$ state. The evolution of the observed populations is shown in Fig. \ref{fig:threeion}.

For the two-ion single-pulse and composite-pulse experiments respectively, we determine the $\ket{{T}}$ state fidelities from histograms obtained with 30,000 and 20,000 separate measurements respectively; when applying the analysis pulses to the $\ket{{T}}$ state, we determine the populations from histograms obtained with 1,500 and 1,000 measurements for each of the conditions $\Phi=\pi\frac{N}{10}$, $N=0-19$.  We determine the $\ket{{W}}$ state populations from histograms obtained with 20,000 separate measurements; when applying the analysis pulses to the $\ket{{W}}$ state, we determine the populations from histograms obtained with 1,000 measurements for each of the conditions $\Phi=\pi\frac{N}{10}$, $N=0-19$.

\begin{figure*}
  \centering
\includegraphics[width=17 cm]{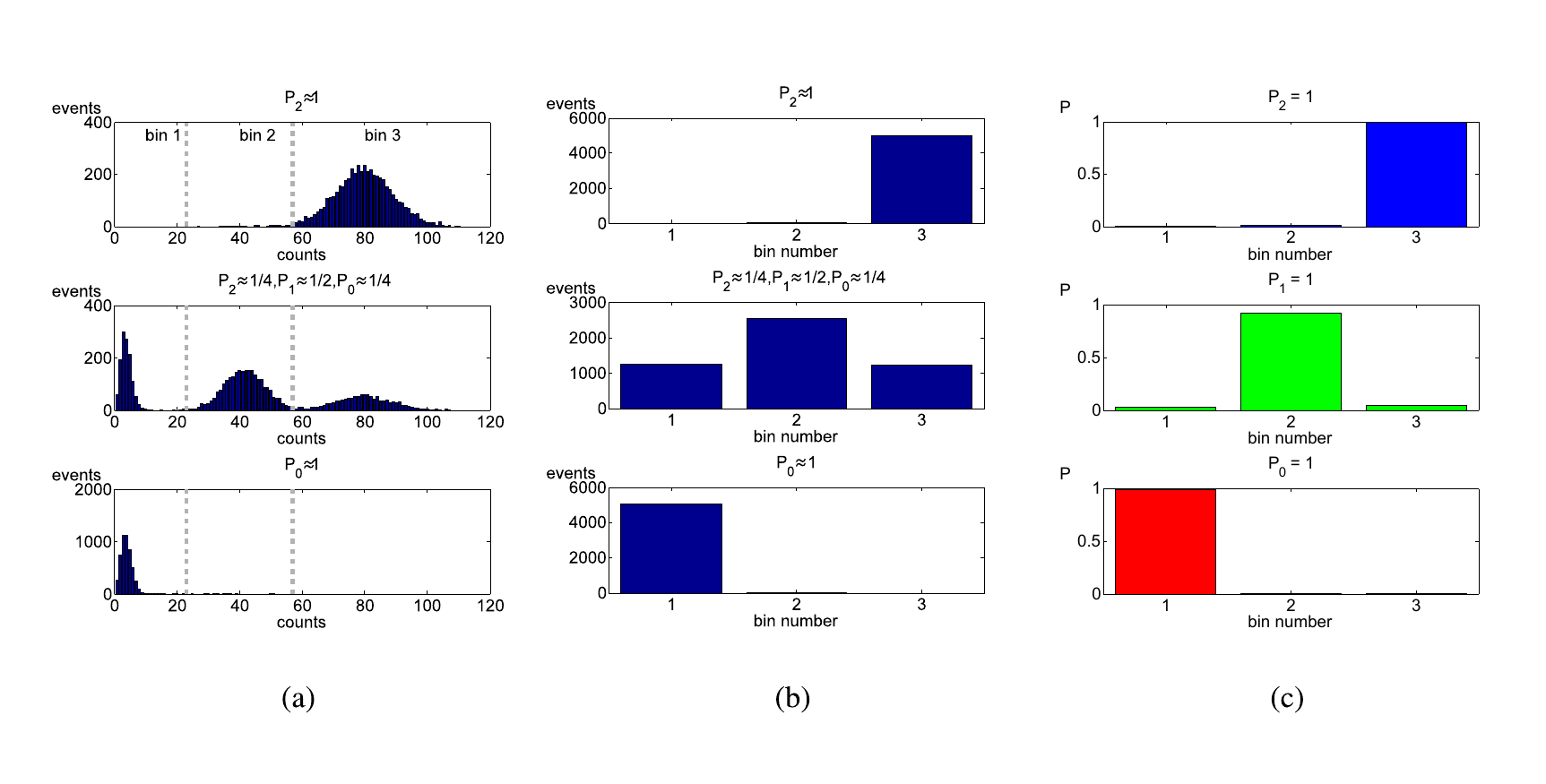}

  \caption{Typical histograms collected after preparation of well-defined state populations for two ions. (a) Three full histograms collected after certain reference pulse sequences. The intended populations are noted on top of each histogram. From top to bottom, histograms for the populations indicated as shown, corresponding to $\Phi=\{0,\pi/2,\pi\}$ (see text). Gray vertical dashed lines separate the histogram into sections of count ranges that can distinguish different numbers of ions in the bright state. $P_i$ denotes $i$ ions in the bright state. (b) Counts within each section in (a) are summed up to form ``rebinned'' histograms. The new bins are labeled as 1-3. (c) From the rebinned histograms and the predicted populations due to the applied rotations, a set of inferred count distributions representing 0-2 ions in the bright state is obtained with a maximum likelihood method as described in the text.
   }  \label{fig:simple_binning}
\end{figure*}

For the ML partial tomography algorithm, we record histograms of an unknown state to be determined as well as states for which we assume the populations are known. The former are called data histograms, the latter reference histograms. To measure spins along different axes of the Bloch sphere, the unknown state can be modified by a known unitary rotation - an analysis pulse - before recording a data histogram as described above. The reference histograms are used to derive count distributions for $n$ ions in the bright state, where $n$ is an integer. These are then used to extract spin populations from the data histograms. For the two-ion experiment, $n$ is 0, 1, or 2; for the three-ion experiment, $n$ ranges from 0 to 3. To obtain the reference histograms, we first optically pump the ions to the $\ket{{2,2}}$ state. We then drive transitions between the $\ket{{2,2}}$ state and the $\ket{{1,1}}$ state by
applying a pulse sequence $\{3\pi/2,0\}$-$\{\pi/2,\Phi\}$ with $\Phi$ sampled from $N\pi/4$, $N=0-7$. The reference histograms are obtained by subsequent fluorescence detection of the laser-induced cycling transition $\ket{{2,2}}\leftrightarrow2p~^2P_{3/2}\ket{{3,3}}$ followed by transferring the $\ket{{1,1}}$ state to the $\ket{{1,-1}}$ and $\ket{{1,0}}$ states via the $\ket{{2,0}}$ state. We repeat the process 6,000 times for each value of $\Phi$ to accumulate photon-counts for reference histograms.  For these initial reference histograms, we restrict ourselves to separable states where we can assume that the state preparation and rotations are of much higher fidelity than the operations to produce entangled states. This assumption was independently verified in separate calibration experiments. For efficiency, we bin several channels of the original histograms together to reduce the number of parameters that need to be inferred. This strategy takes advantage of the fact that the histograms have much more information than is necessary for inferring the parameters of interest with sufficiently low uncertainty.  We bin contiguous ranges of fluorescence counts by means of a heuristic that minimizes loss of information while trying to introduce as few bin boundaries as possible. A simple example of this binning strategy with actual experimental reference histograms is depicted in Fig. \ref{fig:simple_binning}, where we use three bins. For the analysis of actual experimental histograms, we choose five bins for the two-ion experiment and seven bins for the three-ion experiment. We use $10~\%$ of each of the reference histograms exclusively to determine the bin boundaries which are then fixed for analyzing the remainder of the reference data and the histograms of the entangled states. The bin boundaries also remain fixed when extracting the uncertainty estimates.

We can, in principle, use a maximum likelihood method to find the binned count distributions for exactly $n$ ions in the $\ket{{\uparrow}}$ state that optimally match the known spin initialization and microwave rotations, which we assume to be perfect. Subsequently, the inferred count distributions can be used to extract populations in the $\ket{{\uparrow}}$ state for each data histogram. We assume that the reference histograms are consistent with the data and compute likelihoods with a probabilistic model, which simultaneously assigns the probability of observing counts in both the reference histograms and the data histograms of the unknown state. Here, we summarize the method, which will be detailed in a future publication. On one hand, we assign reference probability distributions (Fig. \ref{fig:simple_binning}(c)) and compare to each of the references (Fig. \ref{fig:simple_binning}(b)); on the other hand, under the assumption that the initial state preparation and the analysis pulses are perfect, we can assign a density matrix for the experiment output state and use the reference probability distributions to compare with the data histograms. The joined results of the above two processes lead to an overall likelihood. We then alternate between maximizing the likelihood by varying reference count probabilities with the assigned density matrix fixed and maximizing it by varying the assigned density matrix with the reference count probabilities fixed. The inferred reference count probabilities can be improved by standard techniques for convex optimization over a polytope. For the density matrix, we use the ``$R\rho R$'' algorithm that keeps the estimated density matrix physical while increasing the likelihood at each iteration \cite{Hradil2004}.  Because we do not use an informationally complete set of measurements, the likelihood is maximized equally by any of a set of density matrices that are indistinguishable by our measurements. One of these density matrices, $\rho_{\text{ML}}$, is identified by the ML algorithm. However, because the measurements are sufficient to estimate the experimental states' fidelity with respect to the desired target state $\ket{\psi}$, all possible maximum likelihood density matrices yield the same fidelity $F_{\psi}=Tr(\ket{{ \psi}}\bra{{ \psi}}\rho_{\rm ML})$. This is ensured by the design of the unitary rotations to analyze the states: After applying a rotation $U_i$, we perform fluorescence detection and collect count histograms that correspond to measurements of $n$ ions in the $\ket{{\uparrow}}$ state as $Tr(A_n U_i\rho U_i^\dagger)$, where for example $A_{n=0,1,2}=\{\ket{{\downarrow\downarrow}}\bra{{\downarrow\downarrow}},\ket{{\uparrow\downarrow}}\bra{{\uparrow\downarrow}}+\ket{{\downarrow\uparrow}}\bra{{\downarrow\uparrow}},\ket{{\uparrow\uparrow}}\bra{{\uparrow\uparrow}}\}$, respectively. The $U_i$'s are designed for fidelity measurement of target state $\ket{{\psi}}$ such that there exists a linear combination of $Tr(A_n U_i\rho U_i^\dagger)$ that is equal to the projector onto the target state $\ket{{\psi}}\bra{{\psi}}$, which yields the overlap, or fidelity, of the  experimental density matrix with the targeted state $\ket{{\psi}}$, but not sufficient for the entire density matrix. Thus, the tomography is partial in the sense that not all features of the unknown state are inferable, but the relevant populations and fidelities are.

The uncertainty of inferred quantities such as the fidelities of interest are obtained by parametric bootstrap resampling with 500 resamples \cite{Efron1993}, which determines the 68 \% uncertainty intervals for the fidelities. Since we found that the bootstrap distribution of
the fidelity estimate is approximately symmetric, we estimate a
conservative 68 \% confidence interval for fidelity as
$(F-\epsilon_0-\epsilon_{\rm syst}, F + \epsilon_0)$, where $\epsilon_0=(U -
L)/2$; $U$ and $L$ are the 0.16 and 0.84 quantiles of the bootstrap
distribution respectively and $\epsilon_{\rm syst}$ is a systematic error term (see below). We also computed the log-likelihood-ratios with respect to the
empirical bin frequencies for each of the 500 bootstrapped
analyses, and determined the percentile of the originally found
log-likelihood-ratio in the resulting distribution. This constitutes a
bootstrap likelihood-ratio test for the model used by the analysis
\cite{Oosterhoff1972,Boos2003}. The percentiles found are 22 \% for the two-ion single pulse experiment, 18 \% for the two-ion composite pulse experiment, and 8 \% for the three-ion experiment. These percentiles can be interpreted as bootstrap $p$-value estimates. We also investigate the sensitivity of the inferred entangled state fidelity due to the imperfect initial $\ket{{\uparrow}}$ state preparation. We redo the data analysis assuming the initial density matrices for the reference histograms are $(1-\epsilon)^2\rho_{\uparrow\uparrow}+\epsilon(1-\epsilon)(\rho_{\uparrow\downarrow}+\rho_{\downarrow\uparrow})+\epsilon^2\rho_{\downarrow\downarrow}$ for two ions and
$(1-\epsilon)^3\rho_{\uparrow\uparrow\uparrow}+\epsilon(1-\epsilon)^2(\rho_{\uparrow\uparrow\downarrow}+\rho_{\uparrow\downarrow\uparrow}+\rho_{\downarrow\uparrow\uparrow})+\epsilon^2(1-\epsilon)(\rho_{\downarrow\downarrow\uparrow}+\rho_{\downarrow\uparrow\downarrow}+\rho_{\uparrow\downarrow\downarrow})+\epsilon^3\rho_{\downarrow\downarrow\downarrow}$
for three ions, where $\epsilon$ is the incoherent infidelity per ion. We find that for $\epsilon$ in a range of [0,~0.002], the inferred infidelities are approximately given by $c\epsilon$, where $c$ is a coefficient. We have $\epsilon\leqslant\epsilon_{\rm max}=0.001$ from separate experiments \cite{Gaebler2015}. Thus we obtain an upper bound for a systematic error $\epsilon_{\rm syst}=c\epsilon_{\rm max}$, where the values of $c$ are 0.0021 for the two-ion single pulse experiment, 0.0026 for the two-ion composite pulse experiment, and 0.0025 for the three-ion experiment. Thus we report a conservative 68 \% uncertainty interval for fidelity as $F^{+\epsilon_0}_{-(\epsilon_{\rm syst}+\epsilon_0)}$.

\section{Model for the state evolution with two ions}
For two ions we use a specifically tailored composite pulse sequence. Below we present a theoretical analysis of the two-ion scheme with and without applying the composite pulse sequence.
\subsection{Setup and notation}
\label{SecSetup}
~~~~~The total interaction Hamiltonian $H(t)$ contains a laser-driven sideband coupling $H_s(t)$ and the microwave drive $H_d(t)$,
\begin{align}
H(t) &= H_s(t) + H_d(t)
\label{H}\tag{S5}
\\
H_s(t) &= \hbar \Omega_s (t) (\sigma_1^- - \sigma_2^-) a e^{-i \delta(t) t} + H.c.
\label{Hs}\tag{S6}
\\
H_d(t) &= \hbar\Omega_d(t) \left( \sigma_1^x + \sigma_2^x \right) + H.c.
\label{Hd}\tag{S7}
\end{align}
The Rabi-frequencies $\Omega_s$, $\Omega_d$ and the detuning $\delta$ can be varied in order to maximize the $\ket{{T}}$ state fidelity. In the experiment, we turn on/off the laser beams implementing $H_s$ approximately 0.4 $\rm{\mu s}$, before/after the microwave field implementing $H_s$. In the models discussed here, we only describe the periods when the laser beams and microwave field are acting simultaneously.

In the single-pulse experiment, the time dependence of $\Omega_d(t)$ is given by
\begin{align}
\Omega_d(t) &=
\begin{cases}
\Omega_d, &0 \leq t \leq t_{\pi}
\\
0, &\text{else}
\end{cases}\tag{S8}
\end{align}
where $t_{\pi}$ is the total duration of the pulse.
For the composite pulse scheme discussed in Section \ref{SecEcho}, we assume that the signs of $\Omega_s(t)$ and $\delta(t)$ can be reversed at an intermediate time $0<t_1<t_{\pi}$,
\begin{align}
\Omega_s(t) =
\begin{cases}
+\Omega_s, &t < t_1
\\
-\Omega_s, &t \geq t_1
\end{cases}\tag{S9}
\\
\delta(t) =
\begin{cases}
+\delta, &t < t_1
\\
-\delta, &t \geq t_1.
\end{cases}
\label{Eqdelta}\tag{S10}
\end{align}

\subsection{Entangled state creation}
\label{Section:tpulse}

~~~~~We assume the system is initialized in the state $\ket{{\uparrow\uparrow}}$ (the ions are assumed to be in the motional ground state unless specified otherwise). From this initial state we desire to prepare the triplet state $\ket{T}$ by a single pulse, using the drives $H_d$ and $H_s$ simultaneously.
Rewriting $H_d$ in terms of the states $\ket{{\uparrow\uparrow}}$, $\ket{T}$, and $\ket{{\downarrow\downarrow}}$ yields
\begin{align}
H_{d}(t) = \sqrt{2} \hbar\ \Omega_d(t) \left( \ket{T} \bra{{\uparrow \uparrow}} + \ket{{\downarrow \downarrow}} \bra{T} \right) + H.c.,
\label{Hgate}\tag{S11}
\end{align}
which shows that $H_{d}$ resonantly drives $\ket{{\uparrow\uparrow}}$ to $\ket{T}$ and further on to $\ket{{\downarrow \downarrow}}$.
If the coupling between $\ket{T}$ and $\ket{{\downarrow \downarrow}}$ is turned off, a single pulse of duration $t_{\pi} = \frac{\pi}{2 \sqrt{2} \Omega_d}$ would prepare $\ket{T}$ from $\ket{{\uparrow\uparrow}}$ with unit fidelity. To suppress the coupling to $\ket{{\downarrow \downarrow}}$, we use the sideband coupling $H_s$.\\
~~~~The subspace $S_d$ spanned by $\ket{{\uparrow\uparrow,0}}$ and $\ket{T,0}$ does not couple to the sideband Hamiltonian $H_s$, however, $\ket{{\downarrow \downarrow,0}}$ is coupled to $\ket{S,1}$ and that state is in turn coupled to $\ket{{\uparrow\uparrow, 2}}$ by $H_s$.
These three states form a subspace which we shall refer to as the undesired subspace $S_u$.
As we shall see, the sideband coupling $H_s$ can be engineered to suppress the microwave coupling to the undesired states $\ket{{\downarrow \downarrow,0}}$, $\ket{S,1}$, and $\ket{{\uparrow\uparrow, 2}}$, as discussed in Sec. \ref{SecSubspace}, \ref{SecAnalysis} and \ref{SecSynchronization}. In addition, it is possible to recover population that still leaks to the undesired subspace by the composite pulse technique discussed in Sec. \ref{SecEcho}.

\begin{figure*}[t]
\centering
\includegraphics[width=8.6cm]{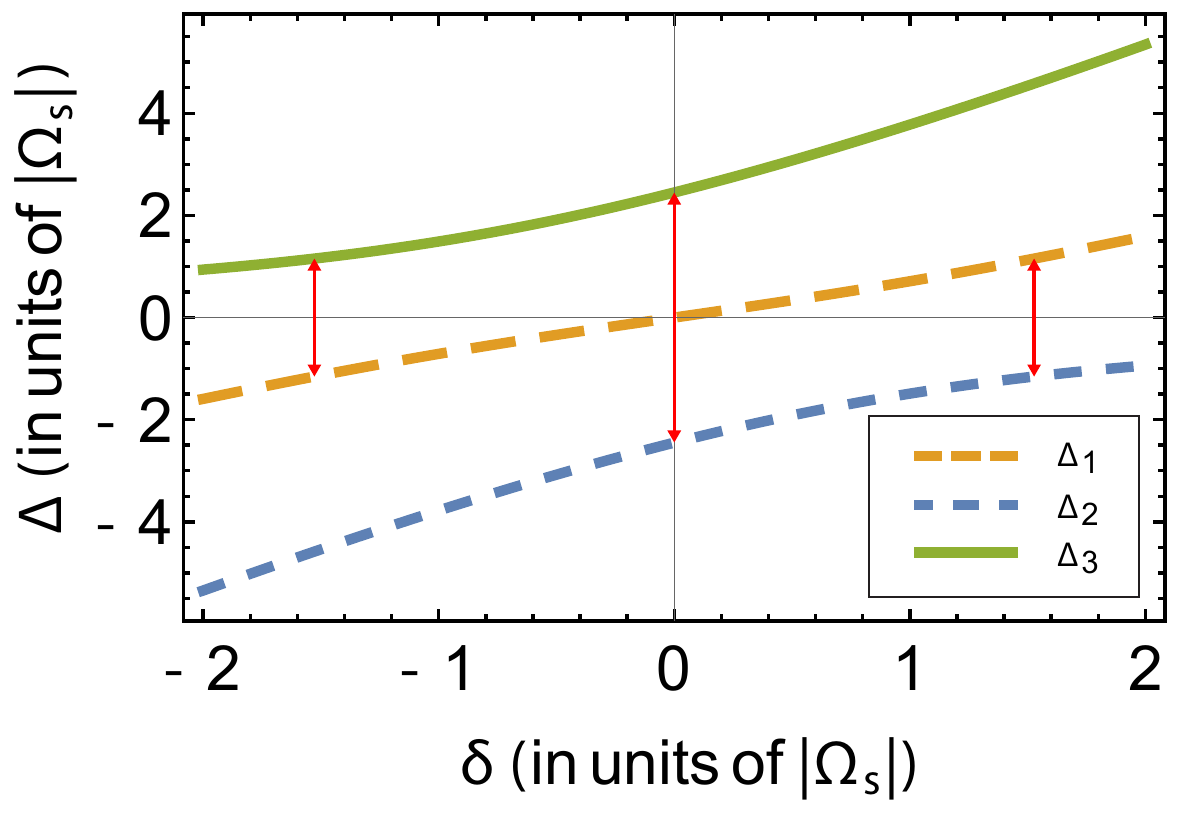}
\caption{
Eigenfrequencies of the coupled subspace consisting of $\ket{{\downarrow\downarrow, 0}}$, $\ket{S, 1}$, and $\ket{{\uparrow\uparrow, 2}}$ as a function of the detuning $\delta$. Values of $\delta$ for which there are two eigenfrequencies of the same absolute value are marked by red arrows. It can be seen that for $\delta = 0$ a dark state with $\Delta = 0$ exists. For $\delta = \pm \sqrt{\frac{7}{3}} \Omega_s$, two of the three eigenfrequencies have the same magnitude and opposite signs, $\Delta_1 = - \Delta_2$, which allows for a commensurate evolution of the dressed states. The third eigenfrequency is $|\Delta_3| \approx 3.97 |\Delta_{1/2}|$ so that we find a close-to harmonic ratio $\Delta_1 : \Delta_2 : \Delta_3 \sim 1:-1:4$.
}
\label{FigEigen}
\end{figure*}

\subsection{Engineering the undesired subspace}
\label{SecSubspace}

We transform the Hamiltonian into a frame rotating with the sideband detuning using a unitary $\mathcal{U}(t) = \exp\left[ i \int_0^t dt' \delta(t') a^\dagger a \right]$ (with $\delta(t)$ as given in Eq. \eqref{Eqdelta}). In this frame we have
\begin{align}
H(t) &= H_\delta(t) + H_s(t) + H_d(t)
\label{Hlab}\tag{S12}
\\
H_\delta(t) &= \hbar\delta(t) ~ a^\dagger a
\label{Ha2}\tag{S13}
\\
H_s(t) &= \hbar \Omega_s(t) (\sigma_1^- - \sigma_2^-) a + H.c.
\label{Hs2}\tag{S14}
\end{align}
and $H_d$ as given by Eq. \eqref{Hd} or \eqref{Hgate} remains unchanged.
Assuming that the motion is initially in the ground state, we can restrict the discussion of the undesired states to $\ket{{\downarrow \downarrow, 0}}$ and the two states coupled to it by the sideband interaction, $\ket{S,1}$ and $\ket{{\uparrow \uparrow,2}}$. The couplings of these states are
\begin{align}
H_{s, \rm u}(t) = \sqrt{2} \hbar \Omega_s(t) \left( \ket{S,1} \bra{{\downarrow \downarrow,0}} - \sqrt{2} \ket{{\uparrow \uparrow,2}} \bra{S,1} \right) + H.c.\tag{S15}
\end{align}
with detunings
\begin{align}
H_{\delta, \rm u}(t) = \hbar \delta(t) \left( \ket{S,1} \bra{S,1} + 2 \ket{{\uparrow \uparrow,2}} \bra{{\uparrow \uparrow,2}} \right).\tag{S16}
\end{align}
It is possible to diagonalize $H_{\rm u} = H_{\delta, \rm u} + H_{s, \rm u}$ and thereby find three dressed states $\ket{\psi_i}$, $i=1,2,3$, of $H_{\rm u}$. However, the expressions for the eigenfrequencies and eigenstates in terms of $\Omega_s$ and $\delta$ are quite involved so that it is difficult to extract conclusions from the expressions. It is therefore helpful to investigate the eigenfrequencies graphically as shown in Fig. \ref{FigEigen}. As will become evident below, it is particularly advantageous to have a harmonic ratio between the eigenfrequencies. We focus on three special cases which are marked with red arrows in Fig. \ref{FigEigen}. For no detuning, $\delta = 0$, we find two eigenfrequencies $\Delta_{1/2} = \pm \sqrt{6} \Omega_s$ which have the same magnitude but opposite signs, and a zero eigenfrequency, $\Delta_0 = 0$, which corresponds to a dark state of $H_{\rm u}$. The zero eigenfrequency makes this parameter choice unattractive, as the dark state would be resonantly coupled to $\ket{T}$. Instead we focus on two other values of $\delta$, where the absolute values of two eigenfrequencies are equal. Since the third eigenfrequency is also nonzero, the microwave couplings from $\ket{T}$ to all three dressed states are nonresonant, and therefore they are only weakly excited. This situation arises at the detuning
\begin{align}\label{eq:delta}\tag{S17}
\delta_{{\rm opt}, \pm} = \pm \sqrt{\frac{7}{3}} \Omega_s,
\end{align}
where we find the eigenfrequencies of $H_{\rm u}$ to be
\begin{align}
\Delta_{1/2} = \pm \frac{2}{\sqrt{3}} \Omega_s, ~ ~ ~ \Delta_3 = + \sqrt{21} \Omega_s ~ ~ ~ \text{(for $\delta_{{\rm opt},+}$)},\tag{S18}
\\
\Delta_{1/2} = \mp \frac{2}{\sqrt{3}} \Omega_s, ~ ~ ~ \Delta_3 = - \sqrt{21} \Omega_s ~ ~ ~ \text{(for $\delta_{{\rm opt},-}$)}.
\label{Eqdressed}\tag{S19}
\end{align}
As we discuss below, when weakly driven the populations of the dressed states will oscillate at their eigenfrequencies. Due to the harmonic ratio $|\Delta_1|:|\Delta_2| = 1:1$ the oscillations of the amplitude on $\ket{\psi_1}$ and $\ket{\psi_2}$ in time will remain synchronized to each other. Quite remarkably, also the ratio of the first two eigenfrequencies and the third eigenfrequency is nearly harmonic, $|\Delta_3| \approx 3.97 |\Delta_{1,2}|$. The amplitude of $\ket{\psi_3}$ thus shares points of nearly the same phase (such as common extrema) with the amplitudes of $\ket{\psi_{1/2}}$ at times which are multiples of $2 \pi / |\Delta_{1/2}|$, provided that the considered time interval is no too long. As we will show in Sec. \ref{SecEcho} we can use these close-to harmonic ratios to ensure that we can find suitable parameters where the amplitudes of all dressed states nearly vanish simultaneously for the composite pulse. Other points where the eigenfrequencies are harmonics of each other, e.g. $|\Delta_1| = 2|\Delta_2|$, exist but are not considered here.

In Sec. \ref{SecAnalysis}, Sec. \ref{SecSynchronization} and Sec. \ref{SecEcho} we will investigate leakage between the subspaces $S_d$ and $S_u$ by means of perturbation theory. To this end, we transform the full Hamiltonian to the dressed state picture using $\delta_{{\rm opt}, +}$ from Eq. \eqref{eq:delta} and the dressed states,
\begin{align}
\ket{\psi_1} &= \sqrt{\frac{3}{118} \left(19-\sqrt{7}\right)} \ket{{\downarrow\downarrow,0}} + \sqrt{\frac{1}{59} \left(19-\sqrt{7}\right)} \ket{S, 1} + \sqrt{\frac{1}{118} \left(23+5 \sqrt{7}\right)} \ket{{\uparrow\uparrow, 2}}\tag{S20}
\\
\ket{\psi_2} &= -\sqrt{\frac{3}{118} \left(19+\sqrt{7}\right)} \ket{{\downarrow\downarrow,0}} + \sqrt{\frac{1}{59} \left(19+\sqrt{7}\right)} \ket{S, 1} + \sqrt{\frac{3}{23+5 \sqrt{7}}} \ket{{\uparrow\uparrow, 2}}\tag{S21}
\\
\ket{\psi_3} &= -\sqrt{\frac{2}{59}} \ket{{\downarrow\downarrow,0}} - \sqrt{\frac{21}{59}} \ket{S, 1} + \frac{6}{\sqrt{59}} \ket{{\uparrow\uparrow, 2}}.\tag{S22}
\end{align}
We thereby obtain the Hamiltonian
\begin{align}
H &= H_0 + H_1
\label{Hbar}\tag{S23}
\\
H_0 &= \hbar \Delta_1 \ket{\psi_1} \bra{\psi_1} + \hbar \Delta_2 \ket{\psi_2} \bra{\psi_2} + \hbar \Delta_3 \ket{\psi_3} \bra{\psi_3} + \hbar \Omega_0 (\ket{T,0} \bra{{\uparrow \uparrow,0}} + H.c.)
\label{H0}\tag{S24}
\\
H_1 &= \hbar \Omega_1 \ket{\psi_1} \bra{T,0}
+ \hbar \Omega_2 \ket{\psi_2} \bra{T,0}
+ \hbar \Omega_3 \ket{\psi_3} \bra{T,0}
+ H.c.
\label{H1}\tag{S25}
\end{align}
Here, we have introduced the shorthand notation for the coupling between $\ket{T}$ and $\ket{{\uparrow \uparrow}}$
\begin{align}
\Omega_0 = \sqrt{2} \ \Omega_d.
\label{Omega0}\tag{S26}
\end{align}
In addition to this desired coupling, the Hamiltonian also contains the undesired couplings from $\ket{T}$ to the three dressed states $\ket{\psi_{1-3}}$ with the coupling strengths
\begin{align}
\Omega_1 = + \sqrt{\frac{3}{59} \left(19-\sqrt{7}\right)} \Omega_d, ~ ~
\Omega_2 = - \sqrt{\frac{3}{59} \left(19+\sqrt{7}\right)} \Omega_d, ~ ~
\Omega_3 = -\frac{2}{\sqrt{59}} \Omega_d,\tag{S27}
\end{align}
and the detunings
\begin{align}
\Delta_1 = +\frac{2}{\sqrt{3}} \Omega_s, ~ ~
\Delta_2 = -\frac{2}{\sqrt{3}} \Omega_s, ~ ~
\Delta_3 = \sqrt{21} \Omega_s.
\label{EqDeltan}\tag{S28}
\end{align}
The coupling from $\ket{T}$ to states other than $\ket{{\uparrow\uparrow}}$ is detuned from resonance and can be suppressed by an arbitrary amount if we assume a sufficiently small ratio $|\Omega_d/\Omega_s| \ll 1$.
In practice, spontaneous emission and other decoherence mechanisms limit how small we can make this ratio, and we need to determine the resulting leakage to the dressed states. We do this by perturbation theory in Sec. \ref{SecAnalysis}.
Secondly, the energies of the dressed states depend only on $\Omega_s$ and $\delta$. The sign of the eigenfrequencies $\Delta_n$ can therefore be reversed by reversing the signs of $\Omega_s$ and $\delta$ simultaneously, while the eigenvectors remain unchanged, a relation that is exploited for the analysis of the composite dynamics in Sec. \ref{SecEcho}, where it is used to ensure destructive interference of the amplitudes on undesired states. We note that while the change $\delta \rightarrow -\delta$ may have some resemblance to spin echo \cite{Hahn1950}, the mechanism that we use is completely different.

\subsection{Dynamics of the microwave excitation}
\label{SecAnalysis}

\subsubsection{Description in perturbation theory}

~~~~~For sufficiently large energy shifts $\Delta_n$ of the dressed states and/or weak enough drive $\Omega_d$ we can use perturbation theory to assess the dynamics of the system.
To this end, $H_0$ in Eqs. \eqref{Hbar}-\eqref{Omega0} is the unperturbed Hamiltonian containing the coupling between $\ket{{\uparrow \uparrow}}$ and $\ket{T}$ in the subspace $S_d$ as well as the undesired states in $S_u$ in the dressed state basis.
Under $H_0$, the subspaces $S_d$ and $S_u$ are uncoupled. The Hamiltonian $H_1$ includes all the couplings between $S_d$ and $S_u$ and will be treated as a perturbation which contains the weak couplings ($\Omega_{1-3} \sim \Omega_d$).
To zeroth order in perturbation theory the initial state $\ket{\psi^{(0)}(0)} = \ket{{\uparrow \uparrow}}$
will evolve into
\begin{align}
\ket{{\psi^{(0)}(t)}} &= c_{\uparrow \uparrow}^{(0)} (t) \ket{{\uparrow \uparrow}} + c_T^{(0)} (t) \ket{T},\tag{S29}
\\
c_{\uparrow \uparrow}^{(0)}(t) &= \cos \left( \Omega_0 t \right),\tag{S30}
\\
c_T^{(0)}(t) &= - i \sin \left( \Omega_0 t \right).\tag{S31}
\label{Eqc10}
\end{align}
A pulse of a duration $t_{\pi} = \frac{\pi}{2 \Omega_0}$ will thus evolve the initial state to $\ket{T}$.
For the excitation of the undesired states we make the ansatz
\begin{align}
\ket{\psi^{(1)}(t)} = \sum_{n=1}^3 c_n^{(1)}(t) \ket{\psi_n}.
\label{Eqc1}\tag{S32}
\end{align}
The dynamics of the coefficients of the dressed states is then described by
\begin{align}
i \dot{c}_n^{(1)} (t) = \Delta_n c_n^{(1)} (t) + \Omega_n c_T^{(0)}(t).\tag{S33}
\end{align}
Solving this for $c_n^{(1)}(0)=0$ yields
\begin{align}
c_n^{(1)}(t) = - \Omega_n e^{- i \Delta_n t} \int_{0}^{t} e^{i \Delta_n t'} c_T^{(0)}(t') dt'.\tag{S34}
\label{Eqc1gen}
\end{align}
Here it is important that $c_T^{(0)}(t) = -i \sin(\Omega_0 t)$ is time-dependent. Solving the integral using the expression for $c_T^{(0)}(t)$ in Eq. \eqref{Eqc10} yields
\begin{align}
c_n^{(1)}(t)
&= \left. - \Omega_n e^{-i \Delta_n t} \frac{e^{i \Delta_n t'}}{\Delta_n^2 - \Omega_0^2} \left(-i \Delta_n \sin \left(\Omega_0 t' \right) + \Omega_0 \cos\left( \Omega_0 t' \right) \right) \right|_0^t\tag{S35}
\\
&= \frac{i \Omega_n}{\Delta_n^2 - \Omega_0^2} \left( \Delta_n \sin \left(\Omega_0 t \right) + i \Omega_0 \left( \cos\left( \Omega_0 t \right) - e^{-i \Delta_n t} \right) \right)
.
\label{Eqc}\tag{S36}
\end{align}
For a weak drive $\Omega_n \ll \Delta_n$, these expressions may be approximated by
\begin{align}
c_{n,{\rm simple}}^{(1)}(t)
\simeq
\frac{i \Omega_n}{\Delta_n} \sin \left(\Omega_0 t \right)
=
- \frac{\Omega_n}{\Delta_n} c_T^{(0)}(t)
.
\label{Eqsimple}\tag{S37}
\end{align}
This expression can be understood by noting that the microwave driving from $\ket{T}$ to $\ket{\psi_{1-3}}$ creates a dressed state $\ket{T'}$ as described by first order perturbation theory
\begin{align}
\ket{T'} = \ket{T} - \sum_n \frac{\Omega_n}{\Delta_n} \ket{\psi_n}.
\label{EqTprime}\tag{S38}
\end{align}
The coefficient in Eq. \eqref{Eqsimple} is seen to contain exactly the same fraction $\Omega_n/\Delta_n$. The term in Eq. \eqref{Eqsimple} thus represents the adiabatic dressing of $\ket{T}$, whereas the remaining terms in Eq. \eqref{Eqc} are diabatic contributions from applying the pulse with a non-vanishing $\Omega_0$.

The population of the dressed state $n$ is given by
\begin{align}
P_n^{(1)}(t) &= |c_n^{(1)}(t)|^2.
\label{EqP}\tag{S39}
\end{align}
Inserting Eq. \eqref{Eqc} gives a rather lengthy expression which is not displayed here.
Keeping only the two leading orders in $\Omega_d / \Omega_s$ we obtain an approximate expression for the population,
\begin{align}
P_n^{(1)} \approx
\frac{\Omega_n^2}{\Delta_n^2} \sin^2\left(\Omega_0 t \right)  - \frac{2 \Omega_0 \Omega_n^2}{\Delta_n^3} \sin\left(\Omega_0 t \right) \sin \left(\Delta_n t \right)
,
\label{EqPapprox}\tag{S40}
\end{align}
from which it can readily be seen that the evolution of the population of the dressed states has two contributions: The first part is the adiabatic part proportional to the population of the $\ket{T}$ state, with an amplitude $\Omega_n^2 / \Delta_n^2$. The second part contains a fast modulation with the frequency $\Delta_n$, at a lower amplitude of $2 \Omega_0 \Omega_n^2 / \Delta_n^2$. We will synchronize these two parts to optimize the protocol in Sec. \ref{SecSynchronization}.

\begin{figure*}[t]
\centering
\includegraphics[width=\columnwidth]{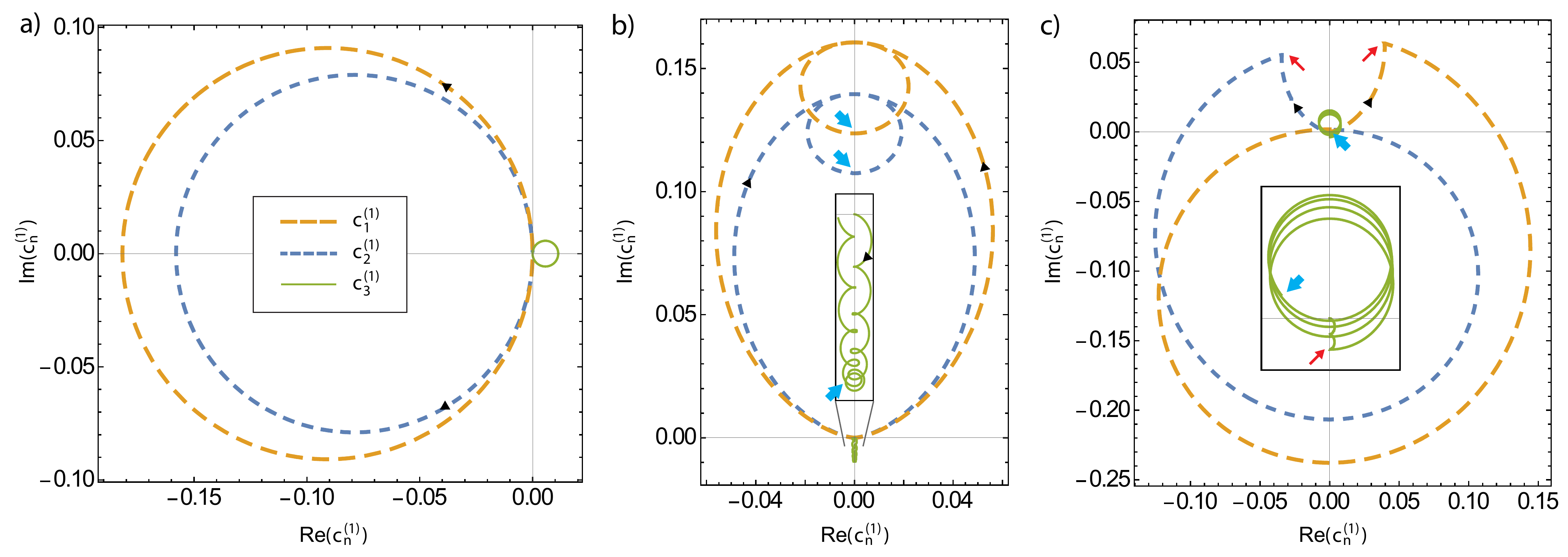}
\caption{
Trajectories of the dressed state coefficients in the complex plane. We plot the evolution of the coefficients $c_n^{(1)}$ of the dressed states $\ket{\psi_{1}}$ (orange long dashed lines), $\ket{\psi_{2}}$ (blue short dashed lines) and $\ket{\psi_{3}}$ (green solid lines) in terms of their real and imaginary part. (a) For $c_T^{(0)}=1$ we find a cyclic evolution which returns to zero at multiples of $t_n = \frac{2 \pi}{|\Delta_n|}$.
(b) For $c_T^{(0)}(t)=-i\sin(\Omega_0 t)$ we find that returning to the origin is only possible around the time where $c_T^{(0)} \approx 0$ and is thus not desirable. Still a minimum of $P_n^{(1)}(t) = |c_n^{(1)}(t)|^2$ can be achieved at $t=t_{\pi}$ for suitable parameters. The corresponding times are marked by thick blue arrows on the dressed state trajectories. In (a) and (b) we use $\Omega_d = 2\Omega_s/(5\sqrt{6})$ corresponding to $m = 1$ in Eq. \eqref{eq:drivingcondition}.
(c) Using a composite pulse technique (described in Sec. \ref{SecEcho}) which reverses the sign of $\Delta_n$ at an intermediate time $t_1 = t_{\pi}/3$ (marked by thin red arrows), it is possible to redirect the trajectories of the dressed states such that the amplitudes vanish at $t=t_{\pi}$ (here shown for $m=1$ in Eq. \eqref{EqOmegaopt} and $\Omega_d=\Omega_s/(3\sqrt{6})$). For the third dressed state $\ket{\psi_3}$ the amplitude nearly vanishes, as can be seen from the insets.
}
\label{FigCoefficients}
\end{figure*}

\subsubsection{State-amplitude evolution}

~~~~~ Before going into details with the optimization, it is instructive to investigate the evolution of the coefficients graphically.
In Fig. \ref{FigCoefficients} we parametrically plot the trajectories of the coefficients $c_n^{(1)}(t)$ in the complex plane for different cases. In a) we first show the evolution assuming time-independent $c_T^{(0)} = 1$. The coefficients move around in circles intersecting at the origin to which they return at multiples of $t_n = 2 \pi / |\Delta_n|$, where $e^{- i \Delta_n t} = 1$. For time-independent $c_T^{(0)}$ and not too large time intervals it would thus always be possible have a very small population in all $\ket{\psi_n}$ at any integer multiple of $t_n$ due to the nearly harmonic ratio $\Delta_1:\Delta_2:\Delta_3 \approx 1:-1:4$.

Taking into account the time dependence of the triplet state coefficient, the evolution in first order perturbation theory, as given by $c_n^{(1)}(t)$ in Eq. \eqref{Eqc} is plotted in b).
The first term in Eq. \eqref{Eqc} moves the coefficient along the imaginary axis. On top of this, the second term in Eq. \eqref{Eqc} represents a combination of a displacement along the real axis and a circular motion. In the limit $\Omega_n \ll \Delta_n$ we are dominated by the displacement along the imaginary axis which is proportional to the triplet state amplitude (since it is mainly caused by the adiabatic dressing as described by Eq. \eqref{EqTprime}). We can thus not find a situation where the amplitude on the dressed state vanishes simultaneously with the population of triplet state being maximal.

\subsection{Harmonic synchronization}
\label{SecSynchronization}

~~~~~In the previous section we found that with constant carrier and sideband driving fields we cannot achieve perfect triplet fidelity. As we will discuss in the following, we can still vary the driving strength $\Omega_d$ to maximize the triplet fidelity by synchronizing the maximum of the triplet population to occur when the dressed state amplitudes go through a local minimum.
\\

From Eq. \eqref{EqPapprox} we see that the temporal evolution of the dressed states populations consists of an envelope with a periodicity in $\Omega_0$ and a modulation with a periodicity in $\Delta_n$. This periodicity gives rise to minimal and maximal values of the population $P_n(t)$ with respect to time. To synchronize a minimum of the oscillations of $P_n(t)$ with the pulse duration, we take $\sin(\Delta_n t_{\pi}) = 1$. For $n=1$ this yields the optimal drive strengths
\begin{align}
\Omega_{d, opt, m} = \frac{|\Delta_1|}{\sqrt{2} (4 m + 1)}, ~ ~ \text{for $m = 0, 1, 2, \dots$}
\label{eq:drivingcondition}\tag{S41}
\end{align}
We can now take advantage of the fact that $|\Delta_1| = |\Delta_2|$ as found for $\delta = \delta_{\rm opt, \pm}$ in Sec. \ref{SecSubspace}. This relation means that we can minimize the population of both $\ket{\psi_1}$ and $\ket{\psi_2}$ simultaneously by the choice in Eq. \eqref{eq:drivingcondition}. In Fig. 7(b) we have used parameters corresponding to $m=1$ and mark the position of the local minimum by the blue arrow. As $\Delta_3$ is much bigger, the population of $\ket{\psi_3}$ is more than an order of magnitude lower and thus less important.
Inserting $\Omega_{d, {\rm opt}, m}$ into Eq. \eqref{Eqc}, we obtain for the coefficients of the dressed states
\begin{align}
c_{n,{\rm opt},m}^{(1)}(t_{\pi})
&= - \frac{i \Omega_n \left( - i |\Delta_n| + \frac{i |\Delta_1|}{4m+1} \right) }{\Delta_n^2- \frac{|\Delta_1|^2}{(4m+1)^2}}
= - \frac{1}{\sqrt{2}(4m+2)} \frac{\Omega_n}{\Omega_d} ~ ~ ~ \text{(for $n = 1, 2$)}\tag{S42}
\\
c_{3,{\rm opt},m}^{(1)}(t_{\pi})
&= - \frac{i \Omega_3 \left( e^{-\frac{i \pi (4 m+1) \Delta_3}{2 \left| \Delta_1 \right| }} \Delta_3 + \frac{i |\Delta_1|}{4m+1} \right) }{\Delta_3^2- \frac{|\Delta_1|^2}{(4m+1)^2}}
\approx - \frac{1}{4 \sqrt{2} (4m+1)} \frac{\Omega_3}{\Omega_d}
.\tag{S43}
\end{align}
In the last step we have assumed $\Delta_3 \approx 4 \Delta_1$ to simplify the expressions. We estimate the infidelity of the protocol by the populations of the undesired states to lowest order. Since the population of $\ket{\psi_3}$ is much smaller than that of $\ket{\psi_{1-2}}$, we can approximate the error of the protocol by only including the first two terms,
\begin{align}
\mathcal{E}_{{\rm opt},m} = 1 - F_{{\rm opt},m} &\approx \sum_{n=1}^2 |c_{n,{\rm opt},m}^{(1)}(t_{\pi})|^2
\approx \frac{1}{4 (1+2m)^2}\tag{S44}
\end{align}
For the three lowest choices of $m$ we obtain the fidelities
\begin{align}
F_{{\rm opt},0} \approx 0.75 ~ ~ ~ \text{(for $m = 0$)}\tag{S45}
\\
F_{{\rm opt},1} \approx 0.97 ~ ~ ~ \text{(for $m = 1$)}\tag{S46}
\label{EqFid1}
\\
F_{{\rm opt},2} \approx 0.99 ~ ~ ~ \text{(for $m = 2$)}\tag{S47}
\label{EqFid2}
\end{align}
In the experiment, we have chosen to operate at $m = 2$. Here we find the optimal driving strength $\Omega_{d,{\rm opt},2} = 2 \Omega_s / (9 \sqrt{6}) \approx \Omega_s / 11$.

\begin{figure*}[!t]
\centering
\includegraphics[width=\columnwidth]{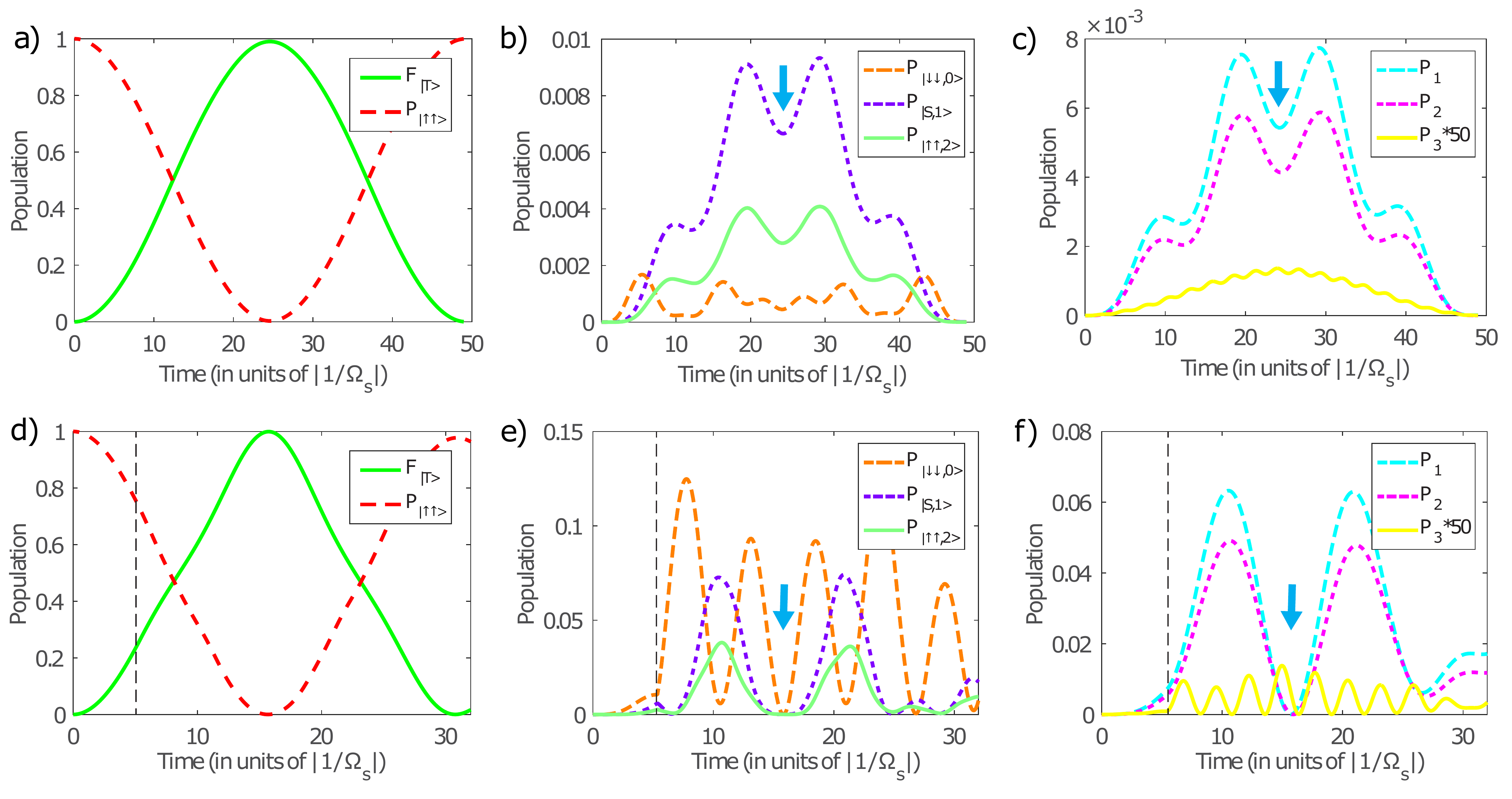}
\caption{Minimization of the population of the undesired states by harmonic synchronization (a)-(c) and composite pulse (d)-(f). (a)-(c) The figures show the result of a numerical simulation of the full Schr\"{o}dinger equation. Adjusting $\Omega_d$ allows for a minimization of the population in the undesired subspace when $F_{\ket{{T}}} $ (green solid line in (a)) is maximized (denoted by thick blue arrows in (b) and (c)). At this point, a half oscillation of $F_{\ket{{T}}} $ contains an integer number of oscillations of the undesired states (b), or of the dressed states resulting from a diagonalization of the undesired subspace (c). The synchronization is possible for the dressed states $\ket{\psi_1}$ and $\ket{\psi_2}$ simultaneously because of the harmonic ratio of their eigenfrequencies, whereas the population of $\ket{\psi_3}$ is considerably smaller. This allows for a fidelity of $F_{\ket{{T}}}  \approx 0.99$ for the experimental parameters ($m = 2$, $\Omega_d/\Omega_s \approx 11$) when neglecting other imperfections. Higher fidelities are achieved with a composite pulse technique (d)-(f) where the signs of $\Omega_s$ and $\delta$ are reversed at a time $t_1$ (denoted by black dashed lines). This makes the minima of the error much smaller, as can be seen from the populations of the undesired states $\ket{{\downarrow \downarrow}}$, $\ket{S, 1}$, and $\ket{{\uparrow \uparrow, 2}}$ in (e) or their dressed states in (f), and can be understood from the trajectories in Fig. \ref{FigCoefficients} (c). In this way, theoretical fidelities $F_{\ket{{T}}}  \gtrsim 0.999$ can be achieved for the experimental parameters ($m = 1$, $\Omega_d/\Omega_s \approx 7$), when neglecting other imperfections.
}
\label{FigPopulations}
\end{figure*}

\begin{figure*}[t]
\centering
\includegraphics[width=\columnwidth]{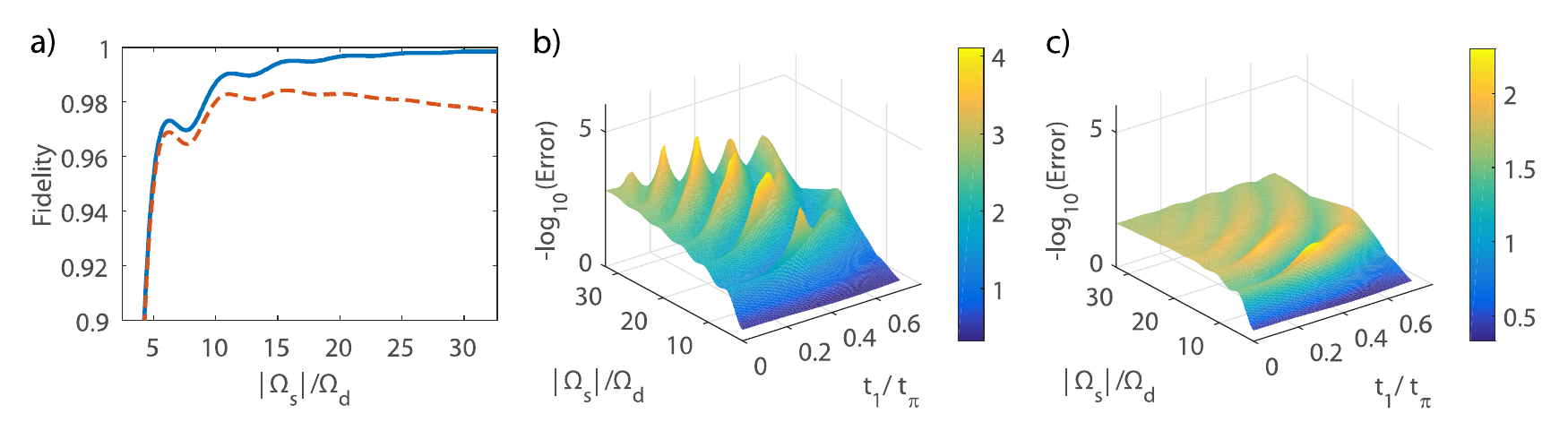}
\caption{Numerically simulated fidelity of the protocols as a function of $\Omega_s/\Omega_d$ and the intermediate time $t_1$. (a) For the single-pulse, the fidelity of the triplet (blue solid line -- without decoherence, red dashed line -- with decoherence) exhibits a periodic behavior as a function of the ratio of the strengths of the sideband and the drive, $\Omega_s/\Omega_{d}$. The maxima are obtained from the harmonic synchronizing condition discussed in Sec. \ref{SecSynchronization}. The first maximum in the figure ($m=1$) is found at $\Omega_s/\Omega_d \approx 6$ and the second ($m=2$) at $\Omega_s/\Omega_d \approx 11$.
In the absence of decoherence, a weaker microwave drive $\Omega_d$ results in a better performance. Larger values of $|\Omega_s|$ increase the infidelity due to spontaneous emission, leading to an optimal drive strength, where both effects are balanced.
(b)-(c) For the composite pulse ($0<t_1<t_{\pi}$), the infidelity (plotted logarithmically) exhibits oscillatory behavior with respect to $\Omega_s/\Omega_d$, and an optimum intermediate time is found at $t_{1, \rm opt} = t_{\pi}/3$, in agreement with the analytical results in Sec. \ref{SecEcho}. This behavior is found both in the absence (b) and in the presence (c) of spontaneous emission. Without spontaneous emission (b), the fidelity increases from $F_{\ket{{T}}}  \approx 0.999$ for the first maximum ($m=1$) at $\Omega_s/\Omega_d \approx 7$ towards higher maxima with lower drive, e.g. $F_{\ket{{T}}}  \approx 0.9999$ at the second maximum ($m = 2$) with $\Omega_s/\Omega_d \approx 14$, whereas with decoherence (c) the fidelity is highest for the first maximum, $F_{\ket{{T}}}  \approx 0.995$, and decreases towards higher ones, e.g. $F_{\ket{{T}}}  \approx 0.991$ for $m=2$.
}
\label{FigPeriodicity}
\end{figure*}

In Fig. \ref{FigPopulations} a)-c) we plot the simulated temporal evolution for $m=2$. Here, as a result of the synchronization, the fast oscillations of the dressed states are symmetric around the maximum of $F_{\ket{{T}}} $. The maximum of the triplet population coincides with a local minimum of the undesired states, as can be seen in the bare (b) and the dressed (c) state pictures of the undesired subspace. In the absence of decoherence, higher fidelities can be achieved for higher $m$, i.e. for more oscillations within the driving pulse. This is shown in Fig. \ref{FigPeriodicity} a), where we plot the fidelity resulting from simulating the Hamiltonian in Eq. \eqref{Hlab} as a function of $\Omega_s / \Omega_d$ with a fixed sideband driving $\Omega_s$ and detuning $\delta$. It can be seen that the optimization by synchronizing the oscillations of the dressed states with the oscillations of $F_{\ket{{T}}} $ allows us to significantly decrease the error.

Performing numerical simulations for the experimental parameters $\Omega_s / \Omega_d \approx 12$, which are close to the derived optimum of $\Omega_s / \Omega_d \approx 11$ for $m=2$, we observe a reduction in the maximal value of the $\ket{T}$ state population of $0.0096$ due to population of states outside the desired subspace in agreement with the result in Eq. \eqref{EqFid2}. Together with infidelities due to other experimental imperfections which we assess below, this imperfection contributes significantly to the experimentally observed infidelity of the single-pulse scheme of $\sim 0.02$. This infidelity can be reduced by making the ratio $\Omega_s / \Omega_d$ larger, which requires reducing the Rabi frequency of the microwaves or increasing the laser power. However, this will lead to increased infidelity due to spontaneous emission, as is discussed in Sec. \ref{SecDecoherence}. Thus, for a given spontaneous emission rate, a compromise emerges from the need to keep $\Omega_d \ll \Omega_s$ and the need to suppress decoherence.
In Fig. \ref{FigPeriodicity} a) we present a simulation including the noise sources discussed below. From this simulation we find that the optimum is around $\Omega_s/\Omega_d=12$ for our conditions, which is the value used in the experiment.
In Fig. 2 in the main text, we plot the populations of the relevant states as predicted by the numerical simulation and find good agreement with the data.

\subsection{Analysis of the composite pulse dynamics}
\label{SecEcho}

~~~~~In the preceding section we have shown how the fidelity can be optimized by synchronizing the oscillations of the dressed states to the envelope of the pulse. We will now show that the attainable fidelity can be further improved by using a composite pulse technique.

In Sec. \ref{SecSetup}, we discussed the possibility to reverse the sign of the sideband coupling, $\Omega_s \rightarrow -\Omega_s$, and the sideband detuning, $\delta \rightarrow -\delta$ at an intermediate time $t_1$. This reverses the Hamiltonian of the undesired subspace $H_{\rm u}$ and thus the energies of the dressed states, $\Delta_n \rightarrow - \Delta_n$, whereas the eigenstates and $\Omega_n$ remain unchanged. This composite pulse sequence allows us to get a cancellation of the population of the dressed states to lowest order in $\Omega_d$. Making the same ansatz as in Eq. \eqref{Eqc1}, the dynamics is described by
\begin{align}
i \dot{c}_n(t) &= +\Delta_n c_n(t) + \Omega_n(t) c_T(t), ~ ~ ~ (t < t_1)\tag{S48}
\\
i \dot{c}_n(t) &= -\Delta_n c_n(t) + \Omega_n(t) c_T(t), ~ ~ ~ (t > t_1).\tag{S49}
\end{align}
The resulting time evolution is given by
\begin{align}
c_{n}^{(1)}(t)
&= - i \Omega_n e^{- i \Delta_n t} \int_{0}^{t} e^{+ i \Delta_n t'} c_T^{(0)}(t') dt', ~ ~ ~ (t < t_1)\tag{S50}
\\
c_{n}^{(1)}(t)
&= c_{n}^{(1)}(t_1) e^{i \Delta_n (t - t_1)} - i \Omega_n e^{+ i \Delta_n t} \int_{t_1}^{t} e^{- i \Delta_n t'} c_T^{(0)}(t') dt', ~ ~ ~ (t > t_1).
\label{Eqcecho}\tag{S51}
\end{align}
The result of the first integral is given in Eq. \eqref{Eqc}. For $t>t_1$ we find
\begin{align}
c^{(1)}_{n}(t)
= - \frac{i\Omega_n}{\Delta_n^2 - \Omega_0^2} &\left[ \Delta_n \left( \sin \left( \Omega_0 t \right) - 2 \sin \left(\Omega_0 t_1 \right) e^{i \Delta_n (t - t_1)} \right) \right. \nonumber
\\
&\left. - i \Omega_0 \left( \cos \left( \Omega_0 t \right) - e^{i \Delta_n (t - 2 t_1)} \right) \right]
.
\label{Eqcecho3}\tag{S52}
\end{align}
For sufficiently weak driving $\Omega_n \ll \Delta_n$, we can derive simplified coefficients to first order in $\Omega_n/\Delta_n$,
\begin{align}
c_{n,{\rm simple}}^{(1)}(t) &\simeq \frac{i \Omega_n}{\Delta_n} \sin\left(\Omega_0 t_1\right) e^{i \Delta_n (t - t_1)} - \left. \frac{i \Omega_n}{\Delta_n} e^{i \Delta_n t} \left( e^{- i \Delta_n t'} \sin\left(\Omega_0 t'\right) \right) \right|_{t_1}^t\tag{S53}
\\
&\simeq
- \frac{i \Omega_n}{\Delta_n} \left( \sin \left(\Omega_0 t \right) - 2 \sin\left(\Omega_0 t_1\right) e^{i \Delta_n (t - t_1)} \right)
.
\label{Eqcechosimple}\tag{S54}
\end{align}
Here the first term which is proportional to the triplet amplitude $c_T^{(0)}(t)$ can again be understood as the adiabatic dressing of the triplet state similar to Eq. \eqref{Eqsimple}. The second term proportional to the triplet amplitude $c_T^{(0)}(t_1)$ at  time $t_1$ is a diabatic contribution resulting from the jump at $t_1$. Because the dressing of $\ket{T}$ is proportional to $\Omega_n/\Delta_n$ (cf. Eq. \eqref{EqTprime}), the change $\Delta_n \rightarrow -\Delta_n$ gives a diabatic contribution by exactly twice the dressing, resulting in the factor of two in Eq. \eqref{Eqcechosimple}.

As opposed to the situation in Sec. \ref{SecSynchronization}, it is now possible to achieve a cancellation $c_n^{(1)}(t_{\pi})=0$ to first order in $\Omega_n/\Delta_n$. This condition is reached if the two terms of the sum interfere destructively, which happens if $\Delta_n(t_{\pi} - t_1) = 2 \pi m$ (for an integer number $m$) and $2\sin(\Omega_0 t_1) = \sin(\Omega_0 t_{\pi}) = 1$. Due to the factor of two in Eq. \eqref{Eqcechosimple} the switching thus has to take place at the time when the triplet state amplitude is $1/2$ which happens at
\begin{align}
t_1 = \frac{t_{\pi}}{3}.\tag{S55}
\label{Eqti}
\end{align}
From the definition of $t_{\pi}=\pi/(2\Omega_0)$, we also obtain a condition on the driving strength $\Omega_d$.
As in the previous section, due to the harmonic ratio of the dressed states $\Delta_1:\Delta_2:\Delta_3=1:-1:\sim 4$, this condition can be fulfilled for dressed states $\ket{\psi_1}$ and $\ket{\psi_2}$ simultaneously, and also approximately for $\ket{\psi_3}$.
With the above conditions, we obtain the optimal driving strength:
\begin{align}
\Omega_{d, {\rm opt}, m} &\approx \frac{|\Delta_{1/2}|}{6 \sqrt{2} m} = \frac{\Omega_s}{3\sqrt{6} m}, ~ ~ ~ m=1,2,...
\label{EqOmegaopt}\tag{S56}
\end{align}

The trajectories in the complex plane corresponding to $m=1$ are shown in Fig. \ref{FigCoefficients} c) and are expected, $c^{(1)}_{1,2}$ are zero at the desired time while $c^{(1)}_3$ are fairly small. The effect of the composite pulse is also evident in Fig. \ref{FigPopulations} d)-f) where we plot the simulated populations of the system states as a function of time, obtained by simulating the Hamiltonian in Eq. \eqref{Hlab}. The populations of the undesired states at the point of maximal triplet population are smaller than in the single-pulse scheme shown in Fig. \ref{FigPopulations} a)-c) and this leads to a significant improvement of the fidelity.
Both in Fig. \ref{FigCoefficients} c) and in Fig. \ref{FigPopulations} d)-f), we use the fastest instance of the composite protocol $m = 1$ where the optimal driving strength is
\begin{align}
\Omega_{d, {\rm opt}, 0} = \frac{\Omega_s}{3 \sqrt{6}} \approx \Omega_s / 7.35,\tag{S57}
\end{align}
which is close to the value of $\Omega_d / \Omega_s \approx 1 / 7$ used in the experiment.

\subsubsection{Second-order dynamics}
\label{SecSecond}

Above we have seen that the excitation of the undesired subspace consisting of the basis states $\ket{{\downarrow\downarrow, 0}}$, $\ket{S, 1}$ and $\ket{{\uparrow\uparrow, 2}}$ or, equivalently, of the corresponding dressed states $\ket{\psi_1}$, $\ket{\psi_2}$, and $\ket{\psi_3}$ can be canceled to first order in $\Omega_d / \Omega_s$. We now consider effects to second order in $\Omega_d/\Omega_s$.
\\

One contribution to second order comes from the terms we neglected when expanding Eq. \eqref{Eqcecho3} to lowest order in Eq. \eqref{Eqcechosimple}.
In addition, the resonant coupling of $\ket{{\uparrow \uparrow, 2}}$ to $\ket{T,2}$ by the microwave drive extends the coupled subspace so that the criteria for which we achieved $c_{1,2}^{(1)} = 0$ are no longer exactly fulfilled. Describing this in full detail is beyond the scope of this analysis. We therefore only estimate the order of magnitude of the error coming from the next order in perturbation theory. As the amplitudes vanish to first order in $\Omega_d/\Omega_s$, the remaining amplitude on undesired components will scale as $\sim \Omega_d^2/\Omega_s^2$ and thus results in a correction to the population of the composite-pulse scheme
\begin{align}
\mathcal{E}^{(2)} &\sim \left(\frac{\Omega_d}{\Omega_s}\right)^4
\overset{m=1}{\approx} 4 \cdot 10^{-4}.
\label{EqErr4}\tag{S58}
\end{align}
This number is consistent with the result of a numerical optimization of the parameters, where we perform simulation of the master equation \eqref{eq:master} using the Hamiltonian in \eqref{Hlab} with no further imperfections. The result of this is shown in Fig. \ref{FigPeriodicity} b)-c).
As can be seen from Fig. \ref{FigPeriodicity} b), for decreasing values of $\Omega_d$, there are many points where the fidelity is very high, corresponding to different $m$ in the above expressions.
Numerically we find an error of $1.2 \cdot 10^{-3}$ when simulating the experimental parameters, $m=1$ and $\Omega_s / \Omega_d \approx 7$. [see also Fig. \ref{FigPopulations} c)-f)]. Fine-tuning of the parameters in the numerical simulation in the vicinity of $m=1$, and setting $t_1=24.18~\rm{\mu s}$ and $t_2=47.57~\rm{\mu s}$, allow for an even smaller error of only $4 \cdot 10^{-4}$ in the absence of additional imperfections, which is consistent with a fourth order contribution. The experiment is, however, dominated by other sources of errors.

\section{Analysis and discussion of experimental imperfections, two-ion case}
\label{SecImperfections}

~~~~In the previous sections, we have seen that with the composite pulse technique it is possible to compensate for leakage to states outside of the desired subspace, the main infidelity of the entangled state generation.
In the following, we provide analysis and discussion of other processes which limit the fidelity of the composite pulse protocol.


\subsection{Spontaneous emission}
\label{SecDecoherence}

~~~~Spontaneous emission through off-resonant excitation of electronically excited states induced by the Raman sideband lasers causes decay from the desired subspace consisting of $\ket{{\uparrow\uparrow}}$ and $\ket{T}$ to other states. We consider the master equation
\begin{align}
\dot{\rho} = -\frac{i}{\hbar} \left[ H(t) , \rho \right] + \sum_k L_k \rho L_k^\dagger - \frac{1}{2} \left( L_k^\dagger L_k \rho + \rho L_k^\dagger L_k \right),
\label{eq:master}\tag{S59}
\end{align}
with Lindblad operators for spontaneous emission
\begin{align}
L_{\downarrow \uparrow, i} &= \sqrt{\gamma_{\downarrow \uparrow}} \ket{{\downarrow}}_i \bra{{\uparrow}}
\label{eq:lindblad1}\tag{S60}
\\
L_{\uparrow \downarrow, i} &= \sqrt{\gamma_{\uparrow \downarrow}} \ket{{\uparrow}}_i \bra{{\downarrow}}
\label{eq:lindblad2}\tag{S61}
\\
L_{o \uparrow, i} &= \sqrt{\gamma_{o \uparrow}} \ket{o}_i \bra{{\uparrow}}
\label{eq:lindblad3}\tag{S62}
\\
L_{o \downarrow, i} &= \sqrt{\gamma_{o \downarrow}} \ket{o}_i \bra{{\downarrow}}
\label{eq:lindblad4}\tag{S63}
,
\end{align}
which model decay processes from level $\ket{{\uparrow}}$ to $\ket{{\downarrow}}$ (Eq. \eqref{eq:lindblad1}) and $\ket{{\downarrow}}$ to $\ket{{\uparrow}}$ (Eq. \eqref{eq:lindblad2}) and from $\ket{{\uparrow}}$ and $\ket{{\downarrow}}$ to a level $\ket{o}$ outside the qubit manifold (Eqs. \eqref{eq:lindblad3} and \eqref{eq:lindblad4}, respectively). The subscripts $i \in \{1,2\}$ denote the ion which undergoes the decay. Here, we have modeled all state outside the qubit space by a single level $\ket{o}$.

The total decay rate out of $\ket{{\uparrow\uparrow}}$ is $\Gamma_{\uparrow \uparrow} = 2 \left( \gamma_{\downarrow \uparrow} + \gamma_{o \uparrow} \right)$ and that from $\ket{T}$ is $\Gamma_{T} = \gamma_{\uparrow \downarrow} + \gamma_{\downarrow \uparrow} + \gamma_{o \uparrow} + \gamma_{o \downarrow}$.
Since $\Gamma_{T}$ and $\Gamma_{\uparrow \uparrow}$ are nearly equal to each other for our parameters, we make the approximations that both $\ket{{\uparrow \uparrow}}$ and $\ket{T}$ decay with the mean decay rate $\bar{\Gamma} = (\Gamma_T + \Gamma_{\uparrow \uparrow})/2$. The reduction of fidelity in $\ket{T}$ due to spontaneous emission is then approximately $P^{(1)}_{\rm spe} (t) \approx 1 - e^{-\bar{\Gamma} t}$, and $\bar{\Gamma}$ is estimated from separate experiments. At the time $t = t_{\pi}$, where the population of $\ket{T}$ is maximal, we obtain $P^{(1)}_{\rm spe} (t_{\pi}) \approx 8 \cdot 10^{-3}$ for the parameters of the single-pulse scheme ($m=2$, $\Omega_s/\Omega_d \approx 12$), and $P^{(1)}_{\rm spe} (t_{\pi}) \approx 5 \cdot 10^{-3}$
for the parameters of the composite scheme ($m=1$, $\Omega_s/\Omega_d \approx 7$), which are consistent with numerically solving the master equation \eqref{eq:master}. In the future, this error can be reduced by tuning the laser frequencies further from the excited states at the cost of a reduced coupling strength, which can be compensated by increased laser power \cite{Ozeri2007}. Another potential solution would be to use a magnetic-field gradient to directly couple the spins to the motion instead of using lasers \cite{Ospelkaus2011,Timoney2011}.

In our analytical calculations in Sec. \ref{SecEcho} we have used the fastest scheme ($m = 1$) for the composite pulse scheme. Numerically, we find by integration of the master equation \eqref{eq:master} that this is indeed the preferred parameter choice for the composite-pulse scheme in the presence of spontaneous emission since longer pulse durations increase the spontaneous emission as shown in Fig. \ref{FigPeriodicity} c).

Assuming spontaneous emission is the only source of incoherent errors, we estimate an error of the $\ket{T}$ state preparation below 0.001 for the single-pulse scheme with a sideband Rabi rate of $2\pi\times17.6$ kHz, a detuning of 29 THz from the $^2P_{1/2}$ state, and a weaker microwave drive with Rabi-frequency $2\pi\times0.23$ kHz. For the composite-pulse scheme with a sideband Rabi rate of $2\pi\times17.3$ kHz, a detuning of 3.8 THz from the $^2P_{1/2}$ state, and a microwave Rabi-frequency of $2\pi\times2.6$ kHz, we can achieve the same error.

\subsection{Imperfect ground-state cooling}


~~~~Imperfect cooling results in a non-zero population of excited motional states, described by their mean occupation number $\bar{n}$, here assumed to be a thermal distribution. In this case, the sideband Hamiltonian Eq. \eqref{Hs2}
perturbs the scheme, as it couples $\ket{{\uparrow\uparrow,n}}$ to $\ket{S,n-1}$, so that $\ket{{\uparrow\uparrow,n>0}}$ is not a dark state of the laser interaction. The coupling from $\ket{{\uparrow \uparrow,n}}$ to $\ket{S,n-1}$ results in the formation of two dressed states at energies $\sim \pm \hbar \sqrt{n} \Omega_s$, whereas $\ket{T,n}$ has energy $0$. The transition from the two dressed states will thus be off-resonant from $\ket{T,n \geq 1}$ so that the $\ket{{T}}$ state preparation is suppressed and nearly all $\ket{{T}}$ state population with $n \geq 1$ is lost from the scheme. This results in an error
\begin{align}
\mathcal{E}_{\bar{n}} \approx \bar{n}. \label{eq:err_nbar}\tag{S64}
\end{align}
In this approximation, the error will be the same both in the single and in the composite pulse case. Using a thermal distribution with $\bar{n} \lesssim 6 \cdot 10^{-3}$ (estimated from sideband ratios after cooling) as the initial density matrix of the motion, we numerically solve the master equation \eqref{eq:master} with the Hamiltonian in Eq. \eqref{Hlab}. From this, we found an infidelity of the $\ket{{T}}$ state due to the imperfect ground-state cooling of less than $6 \cdot 10^{-3}$ for the single-pulse scheme and $5 \cdot 10^{-3}$ for the composite pulse scheme. The upper bound of this error is comparable to that caused by spontaneous emission.

The error due to imperfect ground-state cooling could be reduced with two methods. One method is to tune the Raman sideband laser beams near the second sideband at a frequency difference $\omega_0+2\omega_s\pm\delta^\prime$ such that they only couple states separated by two motional quanta. In this case the scheme will work as long as the motional mode is prepared in either the $\ket{{n=0}}$ or $\ket{{n=1}}$ state, however, the laser interaction strength will be smaller by a factor of the Lamb-Dicke parameter which would lead to slower operation and increased spontaneous emission. An alternative method is to co-trap other species of ions in an ion chain, for example $^9\rm{Be}^+$-$^{25}\rm{Mg}^+$-$^9\rm{Be}^+$ and use the motional mode where the $^{25}\rm{Mg}^+$ ion oscillates out of phase with the $^9\rm{Be}^+$ ions. With the $^{25}\rm{Mg}^+$ ion initialized to  $\ket{{\downarrow_{\rm Mg}}}$, a red sideband coupling $\pi$ pulse driving $\ket{{\downarrow_{\rm Mg},1}}\rightarrow\ket{{\uparrow_{\rm Mg},0}}$ could be applied to the $^{25}\rm{Mg}^+$ ion before the scheme is applied, to increase the probability of initial ground state cooling. Alternatively, after the scheme, the $\ket{{\downarrow_{\rm Mg},1}}\rightarrow\ket{{\uparrow_{\rm Mg},0}}$ pulse could be applied to the $^{25}\rm{Mg}^+$ ion, followed by spin detection on the $^{25}\rm{Mg}^+$ ion. If the $\ket{{\uparrow_{\rm Mg}}}$ state is detected, this gives a partial check that an error occurred during the scheme in which case this preparation sequence could be discarded and we could repeat the preparation process.

\subsection{Ambient heating process}
Ambient heating of the motional mode can lead to additional infidelity. However in the two ion experiment, the motional sideband couples the ions through the stretch mode, which is insensitive to (uniform) electric field noise and has a low heating rate. In general this process is modeled by Lindblad operators of heating and cooling of the motional mode, respectively
$L_{\rm heat} = \sqrt{\gamma_{\rm heat}} a^\dagger$ and $L_{\rm cool} = \sqrt{\gamma_{\rm cool}}a$. We set $\gamma_{\rm cool}=\gamma_{\rm heat}$ which is the observed constant heating rate. This leads to a negligible infidelity for the $\ket{{T}}$ state creation process.

\section{Analysis and discussion of experimental imperfections, three-ion case}

~~~~Using an analysis similar to Sec. 5, for the case of three ions, we determine that, the infidelity of the $\ket{W}$ state has contributions of 0.010 from spontaneous emission, 0.016 from imperfect subspace isolation and less than 0.005 from state preparation. Here, we use the COM mode to prepare the $\ket{W}$ state. This mode experiences a significant ambient heating, approximately 136 quanta/s; simulation indicates this leads to an infidelity of 0.011 for the $\ket{W}$ state preparation. This heating also limits the performance of ground state cooling such that the initial motional state has $\bar{n}\approx0.02$, which leads to an infidelity of 0.018 for the $\ket{W}$ state preparation.

In addition, due to the unequal illumination caused by the finite laser beam waist across the ions, the outer ions experience different AC Stark shifts compared to the center ion, in turn leading to slightly different spin-flip resonance frequencies. From the measured beam waists for the two Raman beams of approximately 28 and 21 $\mu$m, and the 3.35 $\mu$m separation between neighboring ions, we estimate the infidelity from this effect to be 0.023 for a differential AC Stark shift of approximately 2$\pi\times$5 kHz between the center ion and the outer ions. Combining all known effects, the simulation predicts a maximum of the $\ket{W}$ state population of 0.917, in agreement with the experimental result. Since the infidelity for three ions is dominated by other sources than considered in Sec. 4, we expect the gain from using a composite pulse to be small and we do not investigate it here.


\begin{thebibliography}{9}

\bibitem{Nielsen2011}
M.~A. Nielsen, I.~L. Chuang, {\it Quantum Computation and Quantum
  Information\/} (Cambridge University Press, 2011).

\bibitem{Gisin2002}
N.~Gisin, G.~Ribordy, W.~Tittel, H.~Zbinden, {\it Rev. Mod. Phys.\/} {\bf 74},
  145 (2002).

\bibitem{Frerichs1991}
V.~Frerichs, A.~Schenzle, {\it Phys. Rev. A\/} {\bf 44}, 1962 (1991).

\bibitem{Facchi2002}
P.~Facchi, S.~Pascazio, {\it Phys. Rev. Lett.\/} {\bf 89}, 080401 (2002).

\bibitem{Facchi2008}
P.~Facchi, S.~Pascazio, {\it J. Phys. A: Math. Theor.\/} {\bf 41}, 493001
  (2008).

\bibitem{Duer2000}
W.~D{\"u}r, G.~Vidal, J.~I. Cirac, {\it Phys. Rev. A\/} {\bf 62}, 062314
  (2000).

\bibitem{Misra1977}
B.~Misra, E.~C.~G. Sudarshan, {\it J. Math. Phys.\/} {\bf 18}, 756 (1977).

\bibitem{Itano1990}
W.~M. Itano, D.~J. Heinzen, J.~J. Bollinger, D.~J. Wineland, {\it Phys. Rev.
  A\/} {\bf 41}, 2295 (1990).

\bibitem{Balzer2002}
C.~Balzer, {\it et~al.\/}, {\it Opt. Commun.\/} {\bf 211}, 235 (2002).

\bibitem{Maniscalco2008}
S.~Maniscalco, F.~Francica, R.~L. Zaffino, N.~L. Gullo, F.~Plastina, {\it Phys.
  Rev. Lett.\/} {\bf 100}, 090503 (2008).

\bibitem{Wang2008}
X.-B. Wang, J.~Q. You, F.~Nori, {\it Phys. Rev. A\/} {\bf 77}, 062339 (2008).

\bibitem{Raimond2010}
J.-M. Raimond, {\it et~al.\/}, {\it Phys. Rev. Lett.\/} {\bf 105}, 213601
  (2010).

\bibitem{Smerzi2012}
A.~Smerzi, {\it Phys. Rev. Lett.\/} {\bf 109}, 150410 (2012).

\bibitem{Burgarth2013}
D.~Burgarth, {\it et~al.\/}, {\it Phys. Rev. A\/} {\bf 88}, 042107 (2013).

\bibitem{Li2013}
Y.~Li, D.~A. Herrera-Mart{\'\i}, L.~C. Kwek, {\it Phys. Rev. A\/} {\bf 88},
  042321 (2013).

\bibitem{Zhu2014}
B.~Zhu, {\it et~al.\/}, {\it Phys. Rev. Lett.\/} {\bf 112}, 070404 (2014).

\bibitem{Zhang2015}
Y.-R. Zhang, H.~Fan, {\it Sci. Rep.\/} {\bf 5}, 11509 (2015).

\bibitem{Schaefer2014}
F.~Sch{\"a}fer, {\it et~al.\/}, {\it Nature Commun.\/} {\bf 5}, 3194 (2014).

\bibitem{Signoles2014}
A.~Signoles, {\it et~al.\/}, {\it Nature Phys.\/} {\bf 10}, 715 (2014).

\bibitem{Jau2015a}
Y.-Y. Jau, A.~M. Hankin, T.~Keating, I.~H. Deutsch, G.~W. Biedermann, {\it
  Nature Physics\/} {\bf 12}, 71 (2016).

\bibitem{Barontini2015}
G.~Barontini, L.~Hohmann, F.~Haas, J.~Est{\`e}ve, J.~Reichel, {\it Science\/}
  {\bf 349}, 1317 (2015).

\bibitem{Bretheau2015}
L.~Bretheau, P.~Campagne-Ibarcq, E.~Flurin, F.~Mallet, B.~Huard, {\it
  Science\/} {\bf 348}, 776 (2015).

\bibitem{Arecchi1972}
F.~T. Arecchi, E.~Courtens, R.~Gilmore, H.~Thomas, {\it Phys. Rev. A\/} {\bf
  6}, 2211 (1972).

\bibitem{Dicke1954}
R.~H. Dicke, {\it Phys. Rev.\/} {\bf 93}, 99 (1954).

\bibitem{Zeilinger1992}
A.~Zeilinger, M.~A. Horne, D.~M. Greenberger, {\it Workshop on Squeezed States
  and Uncertainty Relations\/} (NASA Conference Publication, 1992), vol. 3135,
  p.~73.

\bibitem{Drees1964}
J.~Drees, W.~Paul, {\it Zeitschrift f{\"u}r Physik\/} {\bf 180}, 340 (1964).

\bibitem{Raizen1992}
M.~G. Raizen, J.~M. Gilligan, J.~C. Bergquist, W.~M. Itano, D.~J. Wineland,
  {\it Phys. Rev. A\/} {\bf 45}, 6493 (1992).

\bibitem{Blakestad2009}
R.~B. Blakestad, {\it et~al.\/}, {\it Phys. Rev. Lett.\/} {\bf 102}, 153002
  (2009).

\bibitem{Langer2005}
C.~Langer, {\it et~al.\/}, {\it Phys. Rev. Lett.\/} {\bf 95}, 060502 (2005).

\bibitem{SUPP}
See supplementary material.

\bibitem{Monroe1995}
C.~Monroe, {\it et~al.\/}, {\it Phys. Rev. Lett.\/} {\bf 75}, 4011 (1995).

\bibitem{Levitt1986}
M.~H. Levitt, {\it Prog. Nucl. Magn. Reson. Spectrosc.\/} {\bf 18}, 61 (1986).

\bibitem{Ozeri2007}
R.~Ozeri, {\it et~al.\/}, {\it Phys. Rev. A\/} {\bf 75}, 042329 (2007).

\bibitem{Reiter2015}
F.~Reiter, {\it et~al.\/}, in preparation.

\bibitem{Haeffner2005}
H.~H{\"a}ffner, {\it et~al.\/}, {\it Nature\/} {\bf 438}, 643 (2005).

\bibitem{Wineland1998}
D.~J. Wineland, {\it et~al.\/}, {\it J. Res. Natl. Inst. Stand. Technol.\/}
  {\bf 103}, 259 (1998).

\bibitem{Hradil2004}
Z.~Hradil, J.~{\v{R}}eh{\'a}{\v{c}}ek, J.~Fiur{\'a}{\v{s}}ek, M.~Je{\v{z}}ek,
  {\it Quantum state estimation\/} (Springer, New York, 2004), pp. 59--100.

\bibitem{Efron1993}
B.~Efron, R.~J. Tibshirani, {\it An introduction to the bootstrap\/} (Chapman
  \& Hall, 1993).

\bibitem{Oosterhoff1972}
J.~Oosterhoff, W.~R. van Zwet, {\it Proceedings of the Sixth Berkeley Symposium
  on Mathematical Statistics and Probability\/} (Univ of California Press,
  1972), vol.~1, pp. 31--50.

\bibitem{Boos2003}
D.~D. Boos, {\it Statistical Science\/} {\bf 18}, 168 (2003).

\bibitem{Gaebler2015}
J.~P. Gaebler, {\it et~al.\/}, in preparation.

\bibitem{Hahn1950}
E.~L. Hahn, {\it Phys. Rev.\/} {\bf 80}, 580 (1950).

\bibitem{Ospelkaus2011}
C.~Ospelkaus, {\it et~al.\/}, {\it Nature\/} {\bf 476}, 181 (2011).

\bibitem{Timoney2011}
N.~Timoney, {\it et~al.\/}, {\it Nature\/} {\bf 476}, 185 (2011).


\end{thebibliography}
\end{document}